\documentclass{pasj01}
\usepackage{url}

\def\uncatcodespecials{\def\do##1{\catcode`##1=12}\dospecials}%
{\catcode`\`=\active\gdef`{\relax\lq}}
\def\setupcode 
    {\tt %
     \spaceskip=0pt \xspaceskip=0pt 
     \catcode`\`=\active
     \uncatcodespecials \obeyspaces
     \catcode`\{=1\catcode`\}=2
    }%
\def\SETUPCODE 
    {\tt %
     \spaceskip=0pt \xspaceskip=0pt 
     \catcode`\`=\active
     \uncatcodespecials \obeyspaces
    }%
\def\docode#1{#1\endgroup}%
\def\code{\begingroup\setupcode\docode}%

\begin{document}
\SetRunningHead{Aihara et al.}{HSC-SSP PDR2}

\title{Second Data Release of the Hyper Suprime-Cam\\Subaru Strategic Program} 

\author{
Hiroaki Aihara\altaffilmark{1,2},
Yusra AlSayyad\altaffilmark{3},
Makoto Ando\altaffilmark{4},
Robert Armstrong\altaffilmark{5},
James Bosch\altaffilmark{3},
Eiichi Egami\altaffilmark{6},
Hisanori Furusawa\altaffilmark{7},
Junko Furusawa\altaffilmark{7},
Andy Goulding\altaffilmark{3},
Yuichi Harikane\altaffilmark{7,8},
Chiaki Hikage\altaffilmark{2},
Paul T.P. Ho\altaffilmark{9},
Bau-Ching Hsieh\altaffilmark{9},
Song Huang\altaffilmark{10},
Hiroyuki Ikeda\altaffilmark{7},
Masatoshi Imanishi\altaffilmark{7},
Kei Ito\altaffilmark{11},
Ikuru Iwata\altaffilmark{7,11},
Anton T. Jaelani\altaffilmark{12},
Ryota Kakuma\altaffilmark{8},
Kojiro Kawana\altaffilmark{1},
Satoshi Kikuta\altaffilmark{11},
Umi Kobayashi\altaffilmark{11},
Michitaro Koike\altaffilmark{7},
Yutaka Komiyama\altaffilmark{7,11},
Xiangchong Li\altaffilmark{4,2},
Yongming Liang\altaffilmark{11},
Yen-Ting Lin\altaffilmark{9},
Wentao Luo\altaffilmark{2,1},
Robert Lupton\altaffilmark{3},
Nate B. Lust\altaffilmark{3},
Lauren A. MacArthur\altaffilmark{3},
Yoshiki Matsuoka\altaffilmark{13},
Sogo Mineo\altaffilmark{7},
Hironao Miyatake\altaffilmark{14,15,2},
Satoshi Miyazaki\altaffilmark{7},
Surhud More\altaffilmark{16},
Ryoma Murata\altaffilmark{1,2},
Shigeru V. Namiki\altaffilmark{11},
Atsushi J. Nishizawa\altaffilmark{14,15},
Masamune Oguri\altaffilmark{17,1,2},
Nobuhiro Okabe\altaffilmark{18},
Sakurako Okamoto\altaffilmark{19},
Yuki Okura\altaffilmark{7},
Yoshiaki Ono\altaffilmark{8},
Masato Onodera\altaffilmark{19},
Masafusa Onoue\altaffilmark{20},
Ken Osato\altaffilmark{21},
Masami Ouchi\altaffilmark{8,2},
Takatoshi Shibuya\altaffilmark{22},
Michael A. Strauss\altaffilmark{3},
Naoshi Sugiyama\altaffilmark{23,2},
Yasushi Suto\altaffilmark{1,17},
Masahiro Takada\altaffilmark{2},
Yuhei Takagi\altaffilmark{19},
Tadafumi Takata\altaffilmark{7},
Satoshi Takita\altaffilmark{7},
Masayuki Tanaka\altaffilmark{7,11,*},
Tsuyoshi Terai\altaffilmark{19},
Yoshiki Toba\altaffilmark{24},
Hisakazu Uchiyama\altaffilmark{7},
Yousuke Utsumi\altaffilmark{25},
Shiang-Yu Wang\altaffilmark{9},
Wenting Wang\altaffilmark{2},
Yoshihiko Yamada\altaffilmark{7},
}
\altaffiltext{1}{Department of Physics, University of Tokyo, Tokyo 113-0033, Japan}
\altaffiltext{2}{Kavli Institute for the Physics and Mathematics of the Universe (Kavli IPMU, WPI), University of Tokyo, Chiba 277-8582, Japan}
\altaffiltext{3}{Department of Astrophysical Sciences, Princeton University, 4 Ivy Lane, Princeton, NJ 08544}
\altaffiltext{4}{Department of Astronomy, Graduate School of Science, The University of Tokyo, 7-3-1 Hongo, Bunkyo, Tokyo, 113-0033, Japan}
\altaffiltext{5}{Lawrence Livermore National Laboratory, Livermore, CA 94551, USA}
\altaffiltext{6}{Steward Observatory, The University of Arizona, 933 North Cherry Avenue, Tucson, AZ 85721-0065, USA}
\altaffiltext{7}{National Astronomical Observatory of Japan, 2-21-1 Osawa, Mitaka, Tokyo 181-8588, Japan}
\altaffiltext{8}{Institute for Cosmic Ray Research, The University of Tokyo, 5-1-5 Kashiwanoha, Kashiwa, Chiba 277-8582, Japan}
\altaffiltext{9}{Institute of Astronomy and Astrophysics, Academia Sinica, 11F of Astronomy-Mathematics Building, AS/NTU No.1, Sec. 4, Roosevelt Rd, Taipei 10617, Taiwan, R.O.C.}
\altaffiltext{10}{Department of Astronomy and Astrophysics, University of California, Santa Cruz, 1156 High Street, Santa Cruz, CA 95064 USA}
\altaffiltext{11}{Department of Astronomy, School of Science, Graduate University for Advanced Studies (SOKENDAI), 2-21-1, Osawa, Mitaka, Tokyo 181-8588, Japan}
\altaffiltext{12}{Department of Physics, Kindai University, 3-4-1 Kowakae, Higashi-Osaka, Osaka 577-8502, Japan}
\altaffiltext{13}{Research Center for Space and Cosmic Evolution, Ehime University, 2-5 Bunkyo-cho, Matsuyama, Ehime 790-8577, Japan}
\altaffiltext{14}{Institute for Advanced Research, Nagoya University, Furocho, Chikusa-ku, Nagoya, 464-8602 Japan}
\altaffiltext{15}{Division of Particle and Astrophysical Science, Graduate School of Science, Nagoya University, Furo-cho, Nagoya 464-8602, Japan}
\altaffiltext{16}{Inter University Centre for Astronomy and Astrophysics, Ganeshkhind, Pune 411007, India}
\altaffiltext{17}{Research Center for the Early Universe, University of Tokyo, Tokyo 113-0033, Japan}
\altaffiltext{18}{Hiroshima Astrophysical Science Center, Hiroshima University, 1-3-1 Kagamiyama, Higashi-Hiroshima, Hiroshima, 739-8526, Japan}
\altaffiltext{19}{Subaru Telescope, National Astronomical Observatory of Japan, 650 N Aohoku Pl, Hilo, HI 96720}
\altaffiltext{20}{Max Planck Institut f\"ur Astronomie, K\"onigstuhl 17, D-69117 Heidelberg, Germany}
\altaffiltext{21}{Institut d'Astrophysique de Paris, Sorbonne Universit\'e, CNRS, UMR 7095, 98 bis boulevard Arago, 75014 Paris, France}
\altaffiltext{22}{Kitami Institute of Technology, 165 Koen-cho, Kitami, Hokkaido 090-8507, Japan}
\altaffiltext{23}{Department of Physics and Astrophysics, Nagoya University, Nagoya 464-8602, Japan}
\altaffiltext{24}{Department of Astronomy, Kyoto University, Kitashirakawa-Oiwake-cho, Sakyo-ku, Kyoto 606-8502, Japan}
\altaffiltext{25}{Kavli Institute for Particle Astrophysics and Cosmology, SLAC National Accelerator Laboratory, Stanford University, 2575 Sand Hill Road, M/S 29, Menlo Park, CA 94025}

\altaffiltext{\ }{\ }
\altaffiltext{*}{\small The corresponding author is Masayuki Tanaka.}
\email{masayuki.tanaka@nao.ac.jp}

\KeyWords{Surveys, Astronomical databases, Galaxies: general, Cosmology: observations}

\maketitle
\definecolor{gray}{rgb}{0.6, 0.6, 0.6}
\newcommand{\commentblue}[1]{\textcolor{blue} {\textbf{#1}}}
\newcommand{\commentred}[1]{\textcolor{red} {\textbf{#1}}}
\newcommand{\commentgray}[1]{\textcolor{gray} {\textbf{#1}}}


\begin{abstract}
  This paper presents the second data release of the Hyper Suprime-Cam Subaru Strategic Program,
  a wide-field optical imaging survey on the 8.2 meter Subaru Telescope.
  The release includes data from 174 nights of observation through January 2018.  
  The Wide layer data cover about 300 deg$^2$ in all five
  broad-band filters ($grizy$) to the nominal survey exposure (10min in $gr$ and 20min in $izy$).
  Partially observed areas are also included
  in the release; about 1100 deg$^2$ is observed in at least one filter and one exposure.  The median seeing in the $i$-band is
  0.6 arcsec, demonstrating the superb image quality of the survey. The Deep (26 deg$^2$) and UltraDeep (4 deg$^2$) data
  are jointly processed and the UltraDeep-COSMOS field reaches an unprecedented depth of $i\sim28$ at
  $5\sigma$ for point sources.  In addition to the broad-bands, narrow-band data are also available in the Deep and UltraDeep fields.
  This release includes a major update to the processing pipeline, including improved sky subtraction, PSF modeling, object detection,
  and artifact rejection.  The overall data quality has been improved, but this release is not without problems; there is
  a persistent deblender problem as well as new issues with masks around bright stars.
  The user is encouraged to review the issue list before utilizing the data for scientific explorations.
  All the image products as well as catalog products are available for download.
  The catalogs are also loaded to a database, which provides an easy interface for users to retrieve data for
  objects of interest.
  In addition to these main data products, detailed galaxy shape measurements  withheld from the Public
  Data Release 1 (PDR1) are now available to the community.  The shape catalog is drawn from
  the S16A internal release, which has a larger area than PDR1 (160 deg$^2$).
  All products are available at the data release site,
  \url{https://hsc-release.mtk.nao.ac.jp/}.
\end{abstract}

\section{Introduction}

Massive imaging and spectroscopic surveys have played an essential role in improving our understanding
of the statistical properties of a wide variety of celestial objects, such as solar system bodies, stars,
galaxies, and active galactic nuclei (AGN).  Surveys are also crucial for modern, high precision cosmology,
and there are a number of ongoing and upcoming surveys that address the nature of dark matter and dark energy.
The Hyper Suprime-Cam Subaru Strategic Program (HSC-SSP; \cite{aihara18b}) is among the most ambitious of the ongoing surveys,
with its aim to cover 1400 deg$^2$ under excellent seeing conditions in multiple filters down to unprecedented depths.

HSC is a wide-field (1.7 degree diameter) optical imager \citep{miyazaki18} installed at the prime focus of the 8.2m Subaru Telescope
operated by National Astronomical Observatory of Japan (NAOJ).  The combination of the field of view and telescope
aperture makes it the most efficient survey instrument to date.
A large imaging survey with this instrument, the HSC-SSP survey, has been awarded 300 nights on the Subaru Telescope; the survey started in March 2014.
The survey consists of 3 layers: Wide, Deep, and UltraDeep.  The Wide layer covers 1400 deg$^2$ in 5 broad-band filters
($grizy$) down to about 26th magnitude.  The Deep layer has 4 separate fields (XMM-LSS, COSMOS, ELAIS-N1, DEEP2-F3)
roughly equally spaced in Right Ascension.
These 4 fields total about 26 deg$^2$.  In addition to the broad-bands, we also observe in 3 narrow-band filters
(NB387, NB816, NB921) in the Deep layer to target emission line objects.
The UltraDeep layer has 2 fields: COSMOS and the Subaru/XMM-Newton Deep Survey (SXDS).
Thanks to long integration times (10-20 hours) in both broad and narrow-bands ($grizy$ and NB816, NB921, NB1010),
we reach to $\sim$28th magnitude over 4 deg$^2$.
For further details, the reader is referred to the survey design paper \citep{aihara18b}.
As of this writing, the survey has used more than 2/3 of the allocated time and has been obtaining excellent data throughout.
Our early science results are summarized in a special issue of the Publications of the Astronomical Society of Japan
(January 2018), which includes exciting results on solar system bodies, stars, galaxies, AGN, and cosmology.

The first public data release (PDR1) from HSC-SSP was made in February 2017, including data taken through
November 2015 from the first 61.5 nights of observations \citep{aihara18a}.  Subsequent incremental releases added
the scientific value of PDR1. The first incremental release happened in June 2017, which
included photometric redshifts for the Wide layer \citep{tanaka18} and deep COSMOS data
from a joint data set taken by the HSC team and astronomers from the University of Hawaii \citep{tanaka17}.  The second
incremental release was in November 2017, including an emission line object catalog \citep{hayashi18},
weak-lensing simulation data \citep{mandelbaum18}, and multi-band SXDS catalog
\citep{mehta18}.  

The current paper presents a new major data release from the HSC-SSP, the second public data release (PDR2).
PDR2 is a major update in terms of both area and depth.  The data quality is also improved thanks to several
important updates made to the processing pipeline.  Thus, PDR2 is a superset of PDR1 in all aspects.  In addition
to the latest survey data, we release the carefully calibrated galaxy shape measurements from \citet{mandelbaum18}
needed for weak-lensing analyses.  All the data products are described in detail in the following sections.

The paper is structured as follows.  We first give a brief summary of PDR2 in Section \ref{sec:the_release}.
Section \ref{sec:hardware_updates} summarizes recent hardware updates, 
followed by a description of improvements to the processing pipeline in Section \ref{sec:pipeline_updates}.
Section \ref{sec:data} describes the data processing as well as a summary of our data products.
Section \ref{sec:data_quality_and_known_issues} presents our data quality assurance tests and
a list of known issues in the release.  A short overview of the data access tools is given
in Section \ref{sec:data_access} and we give an update on our collaborating surveys in Section
\ref{sec:status_of_collaborating_surveys}.   We conclude in Section \ref{sec:summary}
      with a plan for future data releases.

We use the same terminology as in the PDR1 paper to
refer to the data and its processing; see Section 3.1 of \citet{aihara18a} for details.

\section{Overview of the Release}
\label{sec:the_release}

\subsection{The release and changes from PDR1}

\begin{figure*}
 \begin{center}
  \includegraphics[width=18cm]{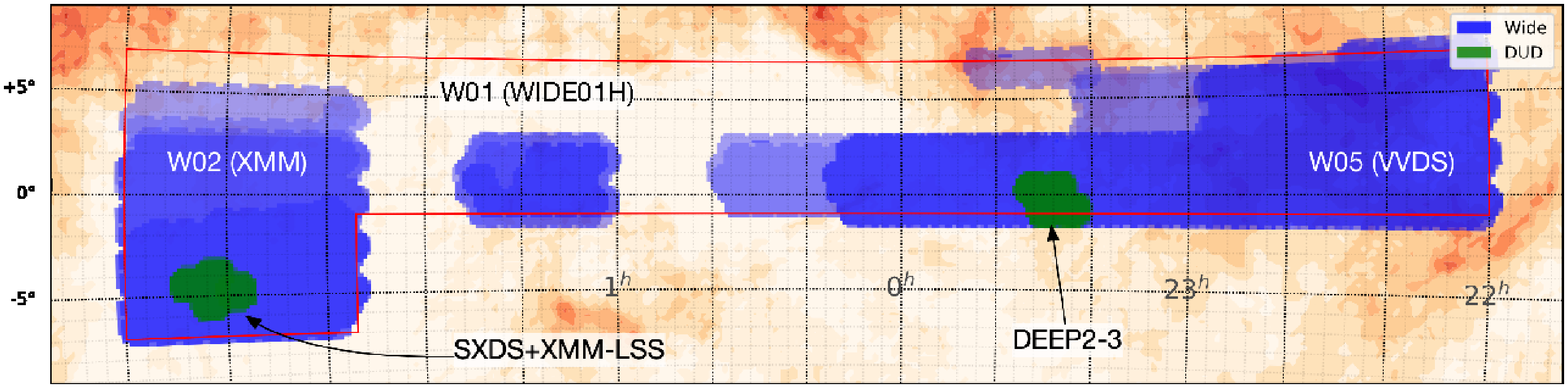}\vspace{1cm}
  \includegraphics[width=18cm]{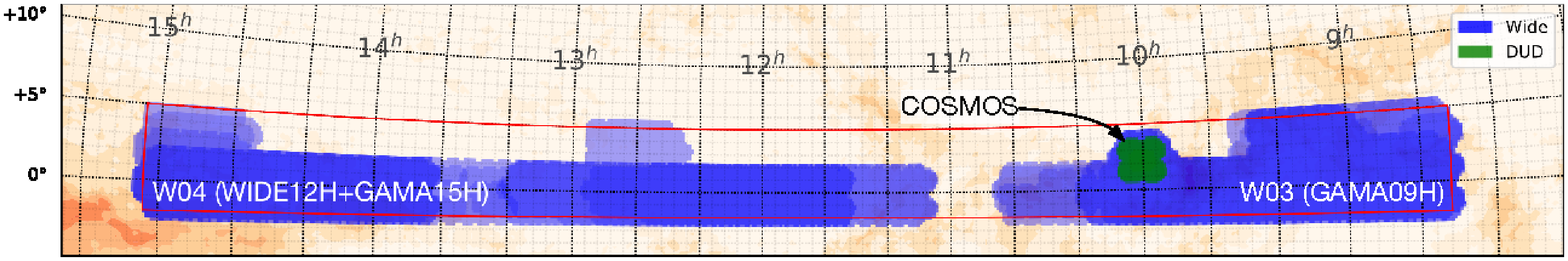}\vspace{1cm}
  \includegraphics[width=18cm]{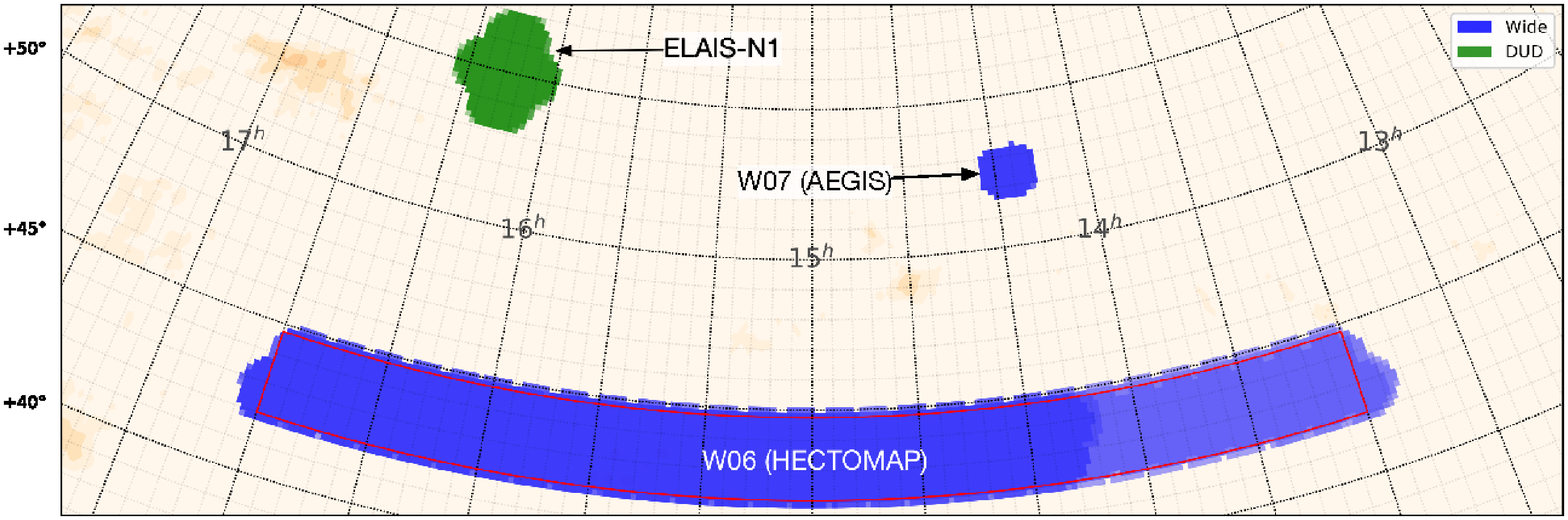} 
 \end{center}
 \caption{
   The area covered in this release shown in equatorial coordinates.
   The blue and green areas show the Wide and Deep+UltraDeep layers, respectively.
   For the Wide layer, the darker color means that the area is observed in more filters (up to 5 filters).
   The red boxes indicate the approximate boundaries of the three disjoint regions that will make up the final Wide survey.
   The Galactic extinction map from \citet{schlegel98} is shown in the background.
 }
 \label{fig:sky_coverage}
\end{figure*}

This release includes data taken from March 2014 through January 2018 from 174 nights of observing time,
including nights lost to weather.  This is a significant increase from the previous release,
which included 61.5 allocated nights.  Fig.~\ref{fig:sky_coverage} shows the survey footprint in PDR2.
Some of the disjoint fields in PDR1 are now connected to each other as the survey has progressed.
Each of the separate Wide layer fields is now given a number; thus they are named W01-W07 as summarized
in Table \ref{tab:field_names}.  Note that the field numbers will change in the next major data release (PDR3)
due to further progress in the survey.

Table \ref{tab:exptime} presents useful global statistics of the data, such
as the exposure time and limiting magnitudes for each filter and survey layer.
Note that the Deep+UltraDeep area is larger in the table than that mentioned in the previous section
because the table includes regions covered in a single exposure (i.e., the area increase is due to dithering).
Major changes since PDR1 include:\\

\begin{itemize}
\item The Wide area which has been observed to the nominal survey depth in all the filters (full-color full-depth
  area in what follows) has increased from about 100 square degrees to 300 square degrees.
\item The Wide layer data in PDR1 included only the full-color full-depth area, but this release
  includes partially observed area as well, i.e., regions not covered in all 5 filters or which have not reached the full depth.
\item The Deep and UltraDeep fields in COSMOS and SXDS overlap each other and they are jointly processed.
  All the coadded images, multi-band catalogs, and database tables are based on the joint data.
\item Due to changes in the processing pipeline, the database table schema have been revised significantly,
  and the table columns have different names, although the correspondence should be obvious in most cases.
  Thus, SQL scripts for PDR1 do not work for PDR2.
\item The $r$ and $i$-band filters are replaced with new and more uniform filters called $r2$ and $i2$, respectively
  (Section \ref{sec:new_filters}).  The release includes data taken with both old and new filters.  They are coadded together.
\item A new sky subtraction algorithm has been implemented that preserves the wings of large objects much
  better than in PDR1 (Section \ref{sec:global_sky_subtraction}).
\item The object detection algorithm has also been improved and the catalog now includes significantly fainter sources
  (Section \ref{sec:dynamic_object_detection}).
\item A new algorithm to remove artifacts from coadds has been introduced.  It works very efficiently in the Wide layer
  (Section \ref{sec:artifact_rejection}).
\item The $y$-band images were significantly affected by scattered light.  This scattered light is now subtracted in
  the processing, resulting in much cleaner $y$-band images (Section \ref{sec:scattered_light_in_the_yband}).
\item A fix to PSFEx, which models the shape of the point-spread function (PSF), has been made, allowing us to
  handle data with very good seeing (Section \ref{sec:psfex_fix}).
  All the good seeing data are used in this release.
\item Lossless compression has been applied to all the pipeline processed images.
  Not all image browsers and I/O interfaces will be able to read these images; the user should use recent
  versions of \code{ds9} and other tools (Section \ref{sec:lossless_image_compression}).
\item Color terms to translate Pan-STARRS1 photometry, which we calibrate our photometry against, have been updated.
  Also, we now exclude late-type stars from photometric calibrations in order to avoid effects of metallicity variations of stars
  (Section \ref{sec:colorterms}).
\end{itemize}

There are, however, known issues in the data, and users are referred to the issue list
in Section \ref{sec:known_issues} before exploring the data for science.
The list is kept up-to-date at the data release website\footnote{\url{https://hsc-release.mtk.nao.ac.jp/}}.

\begin{table*}[htbp]
  \begin{center}
    \begin{tabular}{cccc}
      \hline
      Layer      &  Field Name  &  Database Schema & Database Field Identifier\\
      \hline
      UltraDeep  &  SXDS        &  \code{dud}     & \code{sxds}\\
      UltraDeep  &  COSMOS      &  \code{dud}     & \code{cosmos}\\
      Deep       &  XMM-LSS     &  \code{dud}     & \code{sxds}\\
      Deep       &  E(xtended)-COSMOS & \code{dud}& \code{cosmos}\\
      Deep       &  ELAIS-N1    &  \code{dud}     & \code{elias_n1}\\
      Deep       &  DEEP2-3     &  \code{dud}     & \code{deep2_3}\\
      Wide       &  WIDE01H     &  \code{wide}    & \code{w01}\\
      Wide       &  XMM-LSS     &  \code{wide}    & \code{w02}\\
      Wide       &  GAMA09H     &  \code{wide}    & \code{w03}\\
      Wide       &  WIDE12H     &  \code{wide}    & \code{w04}\\
      Wide       &  GAMA15H     &  \code{wide}    & \code{w04}\\
      Wide       &  VVDS        &  \code{wide}    & \code{w05}\\
      Wide       &  HECTOMAP    &  \code{wide}    & \code{w06}\\
      ---        &  AEGIS       &  \code{wide}    & \code{w07}\\
      \hline 
    \end{tabular}
  \end{center}
  \caption{
    List of the observed fields.  The field names in the Wide layer are left-over from PDR1.
    AEGIS is observed as a photometric redshift calibration field at
    the Wide depth.  The WIDE12H and GAMA15H fields are now connected and they are combined into
    a single field (\code{w04}).  The database field identifier should be used to query for
    a given field.  See the online schema browser for details.  Note that \code{dud} means Deep/UltraDeep.
  }
  \label{tab:field_names}
\end{table*}

\begin{table*}[htbp]
  \begin{center}
    \begin{tabular}{l|ccccccccc}
      \hline\hline
      {\bf Wide}            &  $g$                  &  $r$                 &  $i$                 &  $z$                 &  $y$                 & & & & \\
      exposure (min)        &  $10^{+2}_{-5}$        & $10^{+2}_{-5}$        & $16^{+6}_{-6}$        & $20^{+3}_{-10}$        & $16^{+6}_{-6}$        & & & & \\
      seeing (arcsec)       &  $0.77^{+0.09}_{-0.08}$ & $0.76^{+0.15}_{-0.11}$ & $0.58^{+0.05}_{-0.05}$ & $0.68^{+0.08}_{-0.07}$ & $0.68^{+0.12}_{-0.09}$ & & & & \\
      depth (mag)           &  $26.6^{+0.2}_{-0.3}$   & $26.2^{+0.2}_{-0.3}$  & $26.2^{+0.2}_{-0.4}$   & $25.3^{+0.2}_{-0.3}$   & $24.5^{+0.2}_{-0.3}$   & & & & \\
      saturation (mag)      &  $17.6^{+0.5}_{-0.3}$   & $17.4^{+0.7}_{-0.4}$  & $18.0^{+0.2}_{-0.3}$   & $17.5^{+0.5}_{-0.5}$   & $17.3^{+0.7}_{-0.6}$   & & & & \\
      area (deg$^2$)        &  942                  & 1022                 & 796                  & 905                  & 924                  & & & & \\
      \hline
      {\bf Deep+UltraDeep}  &       $g$            &         $r$          &       $i$            &        $z$           &        $y$           &       $NB387$        &      $NB816$         &      $NB921$         & $NB1010$ \\
      exposure (min)        & $49^{+24}_{-17}$      & $45^{+24}_{-17}$       & $65^{+46}_{-37}$      & $130^{+46}_{-51}$      & $88^{+23}_{-42}$      & $68^{+13}_{-13}$       & $120^{+30}_{-30}$     & $112^{+56}_{-14}$     & ---\\
      seeing (arcsec)       & $0.81^{+0.05}_{-0.13}$ & $0.74^{+0.03}_{-0.05}$ & $0.62^{+0.07}_{-0.07}$ & $0.63^{+0.04}_{-0.03}$ & $0.71^{+0.06}_{-0.06}$ & $0.80^{+0.11}_{-0.08}$ & $0.69^{+0.11}_{-0.12}$ & $0.66^{+0.04}_{-0.07}$ & ---\\
      depth (mag)           & $27.3^{+0.4}_{-0.3}$  & $26.9^{+0.2}_{-0.3}$   & $26.7^{+0.3}_{-0.5}$   & $26.3^{+0.2}_{-0.4}$   & $25.3^{+0.2}_{-0.5}$   & $25.1^{+0.2}_{-0.2}$  & $26.1^{+0.2}_{-0.3}$   & $25.9^{+0.2}_{-0.3}$   & ---\\
      saturation (mag)      & $18.1^{+0.4}_{-0.3}$  & $18.2^{+0.5}_{-0.3}$   & $18.7^{+0.1}_{-0.2}$   & $17.7^{+0.3}_{-0.3}$   & $17.3^{+0.2}_{-0.2}$   & $14.7^{+0.1}_{-0.3}$  & $17.0^{+0.5}_{-0.4}$   & $17.0^{+0.3}_{-0.3}$   & ---\\
      area (deg$^2$)        & 35                   & 35                   & 35                   & 36                   & 36                   & 22                   & 26                   & 28 & ---\\
      \hline \hline
      {\bf Wide}            &       &       &       &       &       &          &          &           & \\
      target exposure (min) &   10  &  10   &   20  &   20  &  20   &          &          &           & \\
      target depth (mag)    & 26.8  &  26.4 &  26.2 &  25.4 &  24.7 &          &          &           & \\
      \hline
      {\bf Deep}            &       &       &       &       &       &          &          &           & \\
      target exposure (min) &   84  &  84   &  126  &  210  & 126   &  84      & 168      &  252      & \\
      target depth (mag)    & 27.8  &  27.4 &  27.1 &  26.6 &  25.6 &  24.8    & 26.1     &  25.9     & \\
      \hline
      {\bf UltraDeep}       &       &       &       &       &       &          &          &           & \\
      target exposure (min) &  420  &  420  &  840  &  1134  & 1134 &          & 630      &  840     & 1050\\
      target depth (mag)    & 28.4  &  28.0 &  27.7 &  27.1  & 26.6 &          & 26.8     &  26.5    & 25.1\\
       \hline \hline
    \end{tabular}
  \end{center}
  \caption{
    Approximate exposure time, seeing, $5\sigma$ depth for point sources,
    and saturation magnitudes (also for point sources) for each filter and survey layer,
    averaged over the entire survey area included in this release.
    The numbers in the top half of the table are the median and the quartiles of the distribution, except for area, which
    shows the total area covered in at least 1 exposure.
    The target exposure times and expected depths (i.e., survey goals) are also shown for reference in
    the bottom half of the table.
    The numbers for the Wide layer shown in the top are close to the full-depth values, while
    those for the Deep+UltraDeep are closer to the Deep depth due to the spatial averaging
    (Deep is wider than UltraDeep).
    Quality assurance (QA) plots showing the depth as a function of position for each field and for each filter
    are available at the data release site.
    Note that the expected depths are for point sources and are in reasonable agreement
    with the measured depths.  The $5\sigma$ limiting mags within 2 arcsec diameter
    apertures, which may be more relevant for extended sources, are shallower by 0.3~mags than
    the point source limits.
    Note that NB1010 is not included in this release.
    Note as well that there is significant spatial variation of all the values listed here over the survey area.
  }
  \label{tab:exptime}
\end{table*}

\subsection{Survey Progress}
\label{sec:survey_progress}

\begin{figure}
  \begin{center}
  \includegraphics[width=9cm]{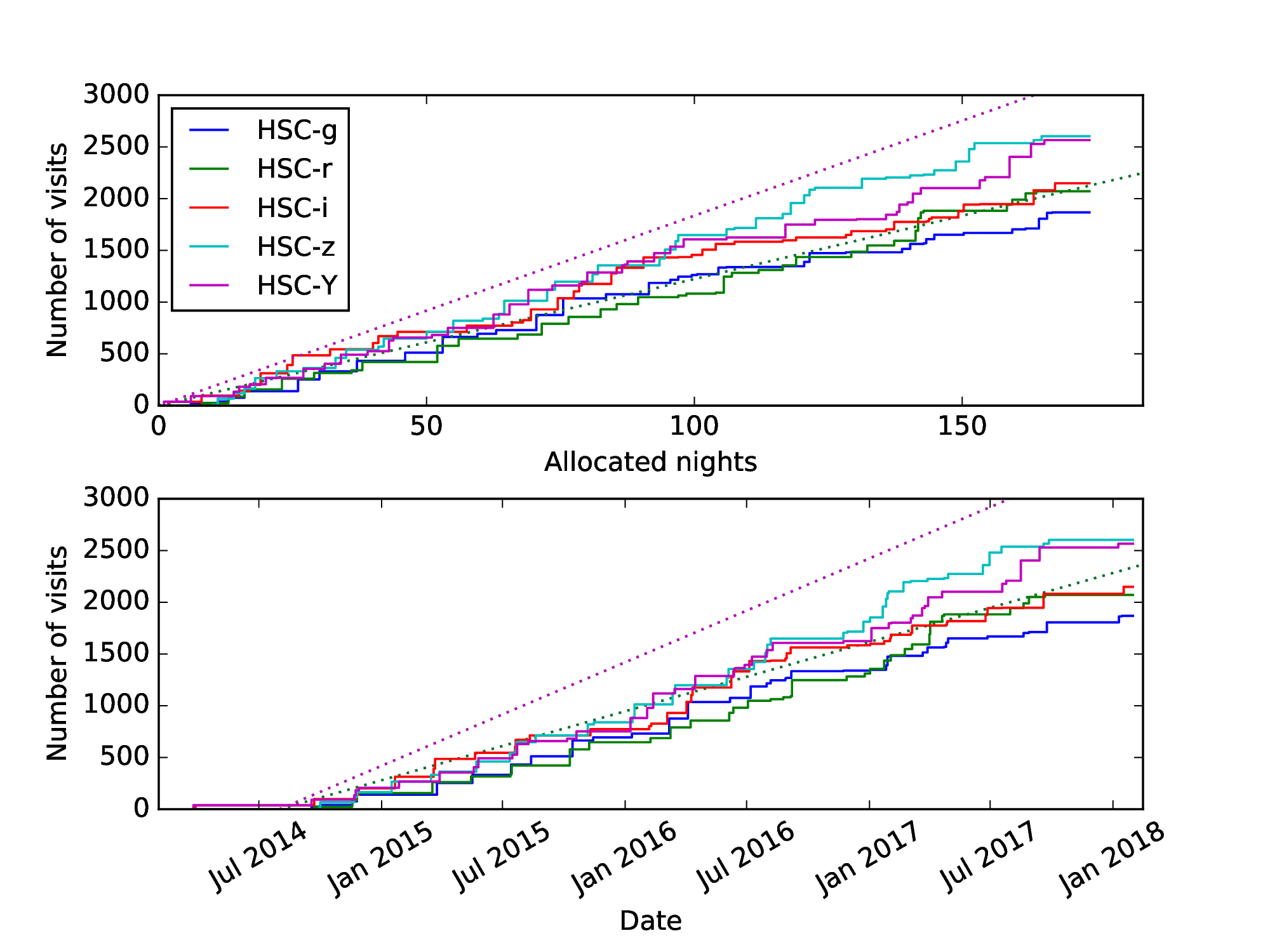} 
  \end{center}
  \caption{
    Allocated number of nights and number of visits acquired for the Wide layer.
    The top panel shows the cumulative number of visits for the Wide layer obtained as
    a function of the number of observing nights.  The dashed lines indicate the average
    numbers of visits required to complete the survey in 300 nights in the $gr$ (bottom line; 4 visits per pointing) and
    $izy$ filters (top line; 6 visits per pointing), respectively.  The bottom panel shows the cumulative number of
    visits as a function of time.  The meanings of the lines are the same as the top panel.
  }
  \label{fig:survey_progress}
\end{figure}

The progress of the Wide survey is summarized in Fig.~\ref{fig:survey_progress}.
This is a good measure of the overall survey progress because two-thirds of
the total observing time is for the Wide survey.  The survey speed has remained
essentially the same since PDR1 (61.5 nights).  The $r$-band is close to
the expected speed, but the other filters are behind schedule.  The completion
rate at the end of January 2018
is 88, 97, 67, 81, and 80\% in the $g$, $r$, $i$, $z$, and $y$ band, respectively.
The $i$-band is the slowest due to the stringent seeing constraint ($\lesssim0.75$ arcsec)
as this is the band in which the weak-lensing analysis is done.
Overall, the survey is progressing at roughly 80\% of the expected speed.  The reason
for the 20\% discrepancy is a combination of optimistic assumptions for overhead between exposures (30 seconds
as opposed to the 20 seconds originally assumed), 30 seconds calibration exposures that
were not included in the original plan, weather, and so on.

\begin{figure}
  \begin{center}
  \includegraphics[width=8cm]{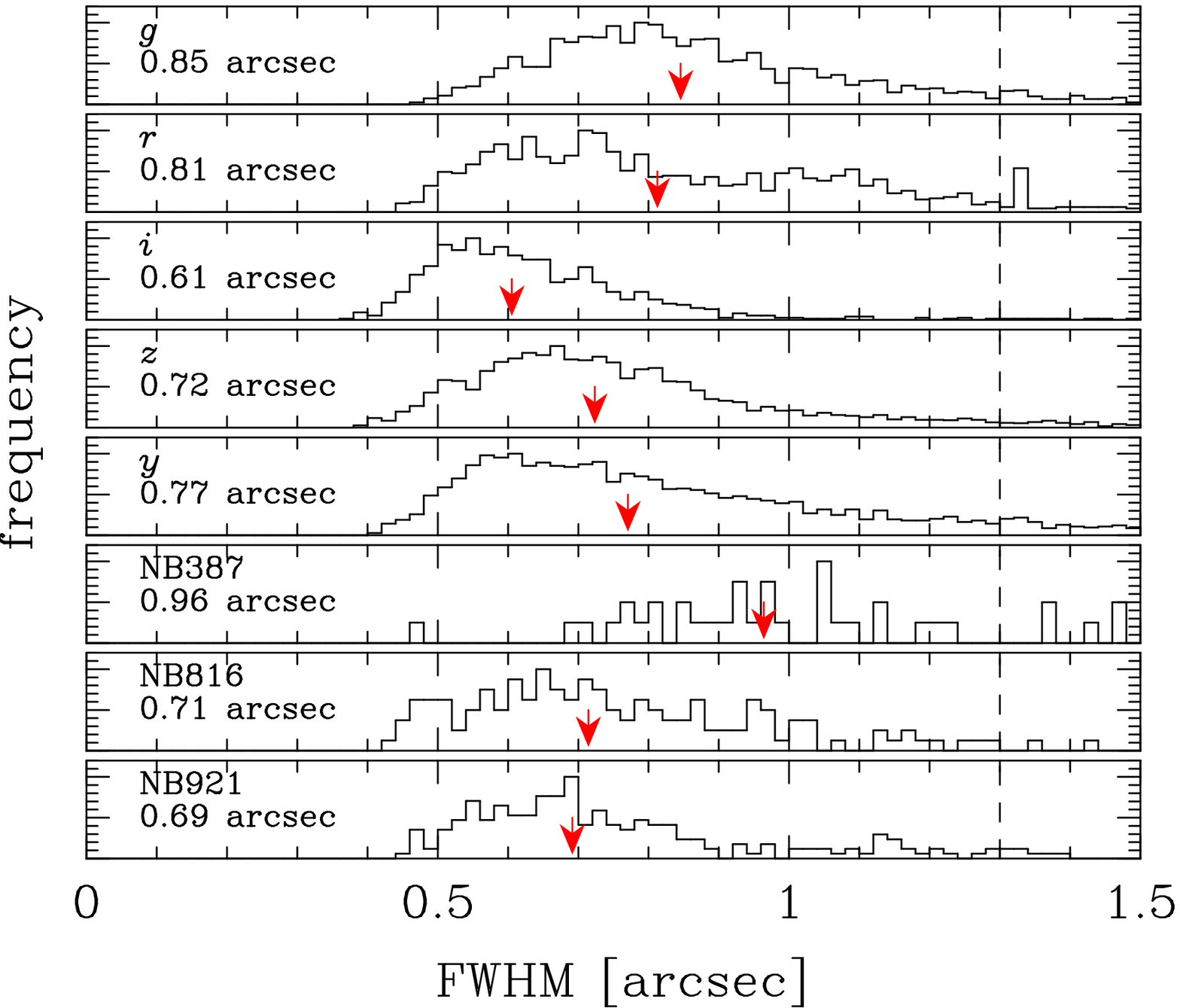} 
  \end{center}
  \caption{
    Seeing distribution of individual visits for each filter.
    The numbers and arrows show the median of the distribution.  The vertical dashed lines indicate
    the seeing threshold (1.3 arcsec) below which visits are used in the processing.
    The plot includes only visits with sky transparency greater than 0.3 (Section \ref{sec:data_screening}).
    Note that seeing shown is as measured and is not corrected for airmass.
  }
  \label{fig:seeing_distrib}
\end{figure}

Nevertheless, the data quality is excellent; Fig.~\ref{fig:seeing_distrib}
shows the distribution of seeing in each visit for each filter; the median $i$-band
seeing is about 0.6 arcsec.  This is superior to other on-going ground-based imaging surveys and is one of
the strengths of the HSC-SSP survey.

\subsection{Previous Internal Releases}
\label{sec:previuos_internal_relesaes}

Our public data releases are based on internal data releases made about 1 year prior to the release.
PDR1 is based on the S15B internal data release, and this PDR2 is based on the S18A data
release made to the HSC collaboration in August 2018.  Table \ref{tab:data_releases} summarizes the
internal releases we have had since PDR1, which have been used in our science papers.
The S16A release was made after S15B and some of our papers in the PASJ special issue published
in January 2018 were based on this release.  S16A was processed with the same pipeline
as in S15B and the data quality remained the same, only the area and depth were increased.
The S17A release incorporated a major pipeline update: the HSC
code branch was merged with the LSST main development stream \citep{juric17}.
The biggest change visible to users was the change in the table schema and the names of various measurement outputs.
However, the correspondence is obvious in most cases.
Finally, the S18A release, on which this PDR2 is based, was made in August 2018 and included
a number of improvements in the data processing algorithms, which we describe in detail in Section \ref{sec:pipeline_updates}.

\begin{table*}[htbp]
  \begin{center}
    \begin{tabular}{l|ccccrrc}
      \hline
      Release               & Date       & Layer          & N      & \multicolumn{1}{c}{Area}        & \multicolumn{1}{c}{Files}    & \multicolumn{1}{c}{N}      & hscPipe \\
                            &            &                & filter & \multicolumn{1}{c}{(deg$^2$)}   & \multicolumn{1}{c}{(TBytes)} & \multicolumn{1}{c}{object} & version \\
      \hline \hline
      Public Data Release 2 & 2019-05-31 & Deep+UltraDeep & 8 &   31        &  88.8 &  20,451,226 & 6.7     \\
      (=S18A)               &            & Wide           & 5 & 1114  (305) & 332.4 & 436,333,410 & 6.7     \\
      \hline \hline
      S17A                  & 2017-09-28 & Deep+UltraDeep & 8 &   31        &  68.0 &  17,506,715 & 5.4     \\
                            &            & Wide           & 5 & 1026  (225) & 209.9 & 348,033,013 & 5.4     \\
      \hline \hline
      S16A                  & 2016-08-04 & UltraDeep      & 7 &    4        &   7.5 &   3,208,918 & 4.0.5   \\
                            &            & Deep           & 7 &   28        &   8.0 &  16,269,129 & 4.0.5   \\
                            &            & Wide           & 5 &  456  (178) & 245.0 & 183,391,488 & 4.0.5   \\
      \hline \hline
      Public Data Release 1 & 2017-02-28 & UltraDeep      & 7 &    4        &   8.6 &   3,225,285 & 4.0.1   \\
      (=part of S15B)               &            & Deep           & 7 &   26        &  16.6 &  15,959,257 & 4.0.1   \\
                            &            & Wide           & 5 &  108  (100) &  57.1 &  52,658,163 & 4.0.1   \\
      \hline
      \hline   
    \end{tabular}
  \end{center}
  \caption{
    Summary of this public release and previous internal data releases. 
    The 5th column gives the survey area covered at least in one filter and one exposure in square degrees.
    The full-color full-depth area in the Wide survey is shown in parentheses.
    The Deep and UltraDeep data have been jointly processed since S17A.
    The 7th column shows the number of primary objects.
  }
  \label{tab:data_releases} 
\end{table*}

\subsection{Calibrated Shape Measurements from PDR1}
\label{sec:calibrated_shape_measurements}

Detailed galaxy shape measurements for weak-lensing analyses were withheld from PDR1.  At this time,
we make these withheld measurements publicly available.  The shape catalog 
we release here is from \citet{mandelbaum18} and is based on the S16A internal data release (see Section  \ref{sec:previuos_internal_relesaes}),
which is larger than PDR1.  
As described in detail in \citet{mandelbaum18}, the catalog covers 136.9~deg$^2$ split into
six separate fields. Galaxy shapes are measured in the $i$-band, in which the mean seeing is $0.58$\arcsec.
Our PSF model meets the accuracy required for
weak lensing science; the fractional PSF size residual is approximately $0.003$ (requirement: $0.004$) and
the PSF model shape correlation function is $\rho_1<3\times 10^{-7}$ (requirement: $4\times 10^{-7}$) at
0.5$^\circ$ scales.  Various null tests are statisticlaly consistent with zero, except for star-galaxy
shape correlations, which reveal additive systematics on $>1^\circ$ scales.
Our first weak-lensing cosmology results presented in \citet{hikage19} are based on this shape catalog.
A number of quality assurance cuts have already been applied (see \cite{mandelbaum18} for details), and
the catalog is ready for science.  It has been loaded to the database under the PDR1 schema,
and the whole set of the S16A data is also included there.

The shape measurements made in PDR2 are withheld for now because they are not fully validated yet. They will be released in the future.
Similarly, deblended images (\code{heavyFootprint}\footnote{
  The processing pipeline attempts to deblend overlapping sources and the result of this process is deblended images,
  which are called \code{heavyFootprint}.
}) are also withheld and will be the subject of a future release.

\section{Hardware Updates}
\label{sec:hardware_updates}

\subsection{New Filters}
\label{sec:new_filters}

The $r$ and $i$ band filters were among the first set of filters manufactured for HSC.
Their filter curves turned out to depend on radius: the cutoff wavelength on the short wavelength side
of the filter transmission curves changes with radial distance from
the filter centers.  The night sky spectrum is very structured  with many strong emission lines.
As the filter bandpass changes with radius, these lines fall in and out of the filter, resulting
in radial structure in the sky background.
This also means that the photometry of detected objects varies across the field of view.  In order to
achieve better photometric accuracy and better background behavior, we manufactured
new filters: $r2$ and $i2$.
These filters were installed on June 24th, 2016 and February 2nd, 2016, respectively.
These new filters have much weaker radial trends; their detailed properties are summarized in \citet{kawanomoto18}.


\subsection{Scattered Light in $y$-band}
\label{sec:encoder_laser_shield}

One of the known issues in PDR1 is that $y$-band images show a pair of arcs due to
scattered light, which were not subtracted very well in the sky subtraction process.
Each of the arcs is about 4-5 arcmin in thickness and crosses the entire field of view.
After some engineering observations, the light source and its path were identified.
The instrument rotator has 8 encoders, each of which has an LED to read a barcode.
Light from the LEDs reflected off the surface of the lens barrel of the Wide-Field Corrector,
thus reaching the detector surface.  The shutter body is wider than the lens barrel, but
is not sufficiently wide to block the oblique incident light.
The scattered light caused a pair of arc-like structures that moved as the instrument rotated.
The exact wavelength of the LED light is unclear, but observations have shown that
the scattered light was seen only in the $y$-band and a few narrow-bands around $0.9-1.0\mu m$.
In order to eliminate the scattered light, covering screens were installed at
the edge of the shutter body on November 13, 2017 to obstruct the light path.
Data taken after that date do not exhibit any sign of scattered light.
However, all data taken before that date were affected, and we have developed software to subtract
the scattered light (Section \ref{sec:scattered_light_in_the_yband}).

\section{Pipeline Updates}
\label{sec:pipeline_updates}

PDR1 was processed with \code{hscPipe v4} as described in detail in \citet{bosch18}.
As mentioned earlier, there have been major pipeline updates since then, and PDR2 is processed
with \code{hscPipe v6}.  This section describes the new features of \code{v6}.

\subsection{Global Sky Subtraction}
\label{sec:global_sky_subtraction}

In previous versions of the pipeline,
background subtraction was performed on each CCD individually.
We used an empirical background model consisting of ``superpixels''
typically of size $256\times 256$ pixels ($43''\times43''$).
A robust measure of the background was obtained for each superpixel by taking
a clipped mean and ignoring \code{DETECTED} pixels, and the superpixels were fit
with a 6th order two-dimensional Chebyshev polynomial.
The superpixels were then interpolated at the regular pixel positions using Akima spline,
and the resultant background image was subtracted from the CCD image.

While simple to implement,
this algorithm has two important drawbacks:
\begin{itemize}
\item The superpixel scale is necessarily limited in size to less than the size of the CCD,
  which means that bright extended objects
  (e.g., nearby galaxies or bright stars)
  can easily bias the sky model.
\item Because CCDs are treated individually,
  there can be discontinuities in the sky model between neighboring CCDs.
\end{itemize}
In order to address these deficiencies,
we developed a new algorithm to perform background subtraction over the entire field-of-view.

The new algorithm incorporates two elements.
The first element is an empirical background model extending over the entire focal plane.
This uses the superpixel technique we used before,
but extends it so that the model can be constructed over the entire focal plane.
Because this model operates across CCD boundaries,
discontinuities at CCD edges are reduced.
Experiments indicate that an appropriate superpixel scale for HSC is $1024\times 1024$ pixels
($\sim2'.8\times2'.8$).
Scales significantly larger leave sky subtraction residuals that vary from exposure to exposure.

\begin{figure}
  \begin{center}
    \includegraphics[width=8cm]{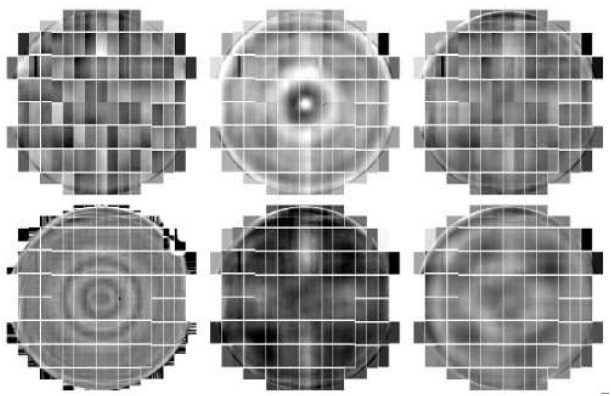}
  \end{center}
  \caption{
    HSC sky frames showing the full focal plane in the  $g$, $r$, $r2$, $i$, $i2$, $z$ bands from top-left to bottom-right.
    The rings in the $r$ and $i$ bands are due to radial variations in the filter curves.
    The $g$ and $r2$ bands show some CCD-dependent features, likely due to CCD sensitivity variations in the blue.
    The other filters not shown here (i.e., $y$ and the narrow-band filters) do not exhibit any significant spatial structure.
  }
  \label{fig:skyFrames}
\end{figure}

The second element is a ``sky frame'',
which is the mean response of the instrument to the sky for a particular filter.
It is constructed from a clipped-mean of the superpixels with objects masked out from many
observations (typically several tens) that have large dithers, so that the same objects do not land on
the same pixels.  This allows subtraction of static features that have a smaller scale
than the empirical background model.   We use superpixels of $256\times 256$ pixels,
which is sufficient to model the ``rings'' in the $r$ and $i$-bands, which are due
to variations in the filter transmission curves as a function of radius from the center
(Section \ref{sec:new_filters}; the rings are essentially gone in the new $r2$ and $i2$ filters).
Fig.~\ref{fig:skyFrames} shows the sky frames.  It is interesting that each filter has
its own characteristic spatial structure.  The systematic offsets between the CCDs seen in blue filters
($g$ in particular) are likely due to variations in the CCD responses, while large-scale patterns
are due to variations in the filter response.

\begin{figure}
  \begin{center}
    \includegraphics[width=4cm]{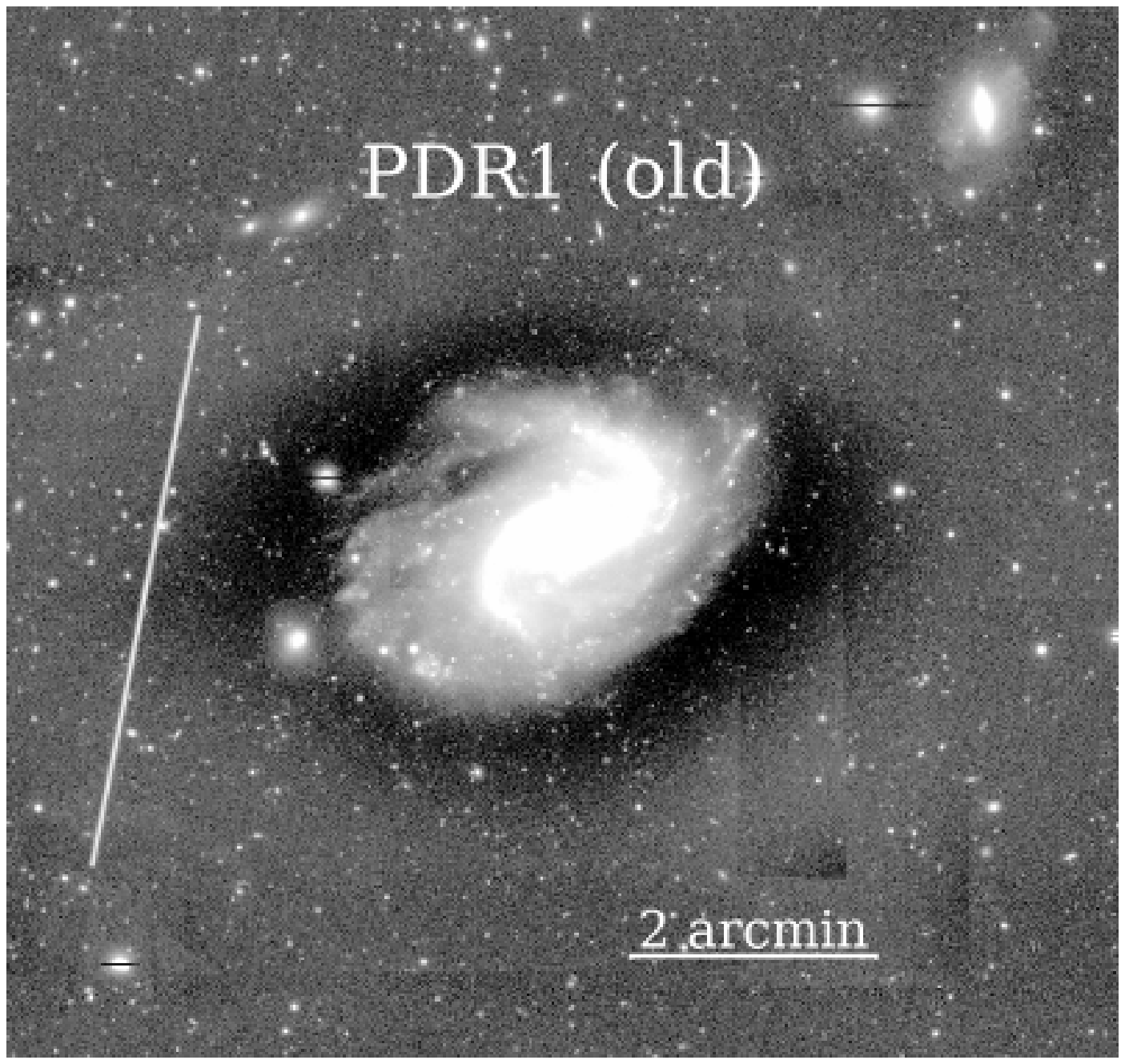}
    \includegraphics[width=4cm]{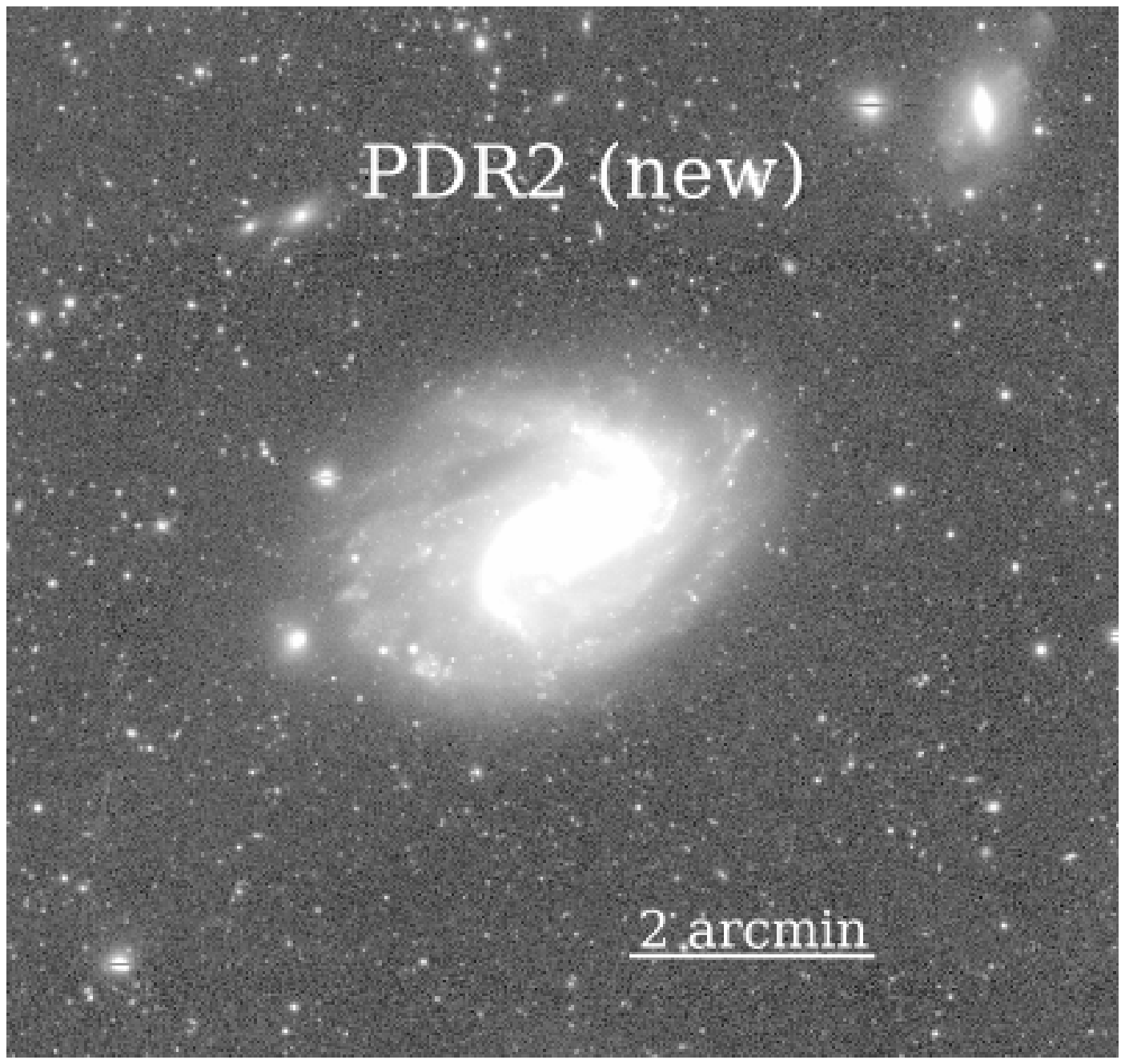}
  \end{center}
  \caption{
    \textbf{Left:} coadd image of a nearby galaxy in the $i$-band from PDR1.
    \textbf{Right:} same image but constructed using the new sky subtraction algorithm.
    The images are stretched to the same level for a fair comparison.
  }
  \label{fig:skySubtraction}
\end{figure}

When subtracting the sky from a science exposure,
we first measure and subtract the large-scale empirical background model (first element),
and then fit and subtract a scaled sky frame (second element).
Fig.~\ref{fig:skySubtraction} compares the old and new algorithms.
The old algorithm (left) tends to subtract extended halos of bright objects as indicated by
the dark halo around the large object at the center.  This has indeed been problematic for
studying nearby galaxies.  The new algorithm (right) preserves the extended
wings much better, demonstrating the improved performance of the sky subtraction.
This improvement is particularly important for large extended sources such as nearby galaxies.
However, masks around bright stars to indicate regions that suffer from false detections and poor photometry
were not revised accordingly
as we discuss in detail in Section \ref{sec:bright_star_masks}.  They will be fixed in
a future incremental release, which will be made by September 1st, 2019.

\subsection{Dynamic Object Detection}
\label{sec:dynamic_object_detection}

In previous versions of the pipeline,
the detection threshold was set statically
as a particular multiple of the noise ($5\sigma$, to be specific).
On coadds, especially coadds with many exposures,
we found that sources that were obviously present in the image were not detected.
We attribute this to an incorrect noise model:
the pipeline tracks an estimate of the variance of each image,
but that estimate can be wrong after convolution operations
since they move a fraction of the variance into covariance,
which is not tracked by the pipeline.

In order to deal with this,
we now set the detection threshold dynamically.
We measure the PSF fluxes for a sample of points chosen to be on empty sky,
avoiding object footprints.  If the variance is perfectly correct,
the standard deviation of the PSF fluxes should agree with the uncertainty
expected from the variance over the effective area of the PSF.
The variance image is not perfect and the ratio between the standard deviation of
these PSF fluxes and the mean of quoted errors provides a correction factor to
the detection threshold.

\begin{figure}
  \begin{center}
    \includegraphics[width=8cm]{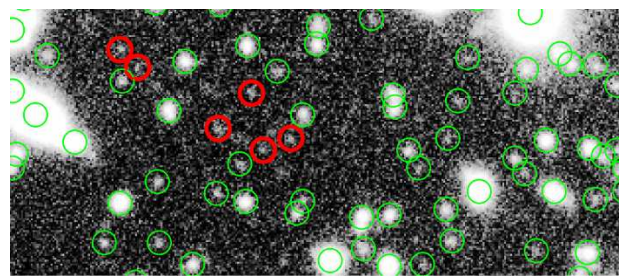}
  \end{center}
  \caption{
    $i$-band image of a small piece of the UltraDeep-COSMOS field ($0'.9\times0'.4$),
    showing the improvement in the detection depth
    using the new dynamic detection feature.
    The green circles indicate sources detected with the old detection algorithm,
    while the red circles indicate sources newly detected with the dynamic detection algorithm.
    The faintest objects in this image are about $27.5-28.0$ mag.
  }
  \label{fig:dynamicDetection}
\end{figure}

Figure~\ref{fig:dynamicDetection} shows an example field
with and without this feature.
There are many faint sources that are missed by the previous detection algorithm
but are detected in the revised algorithm, demonstrating the improvement in
the object detection.  This improvement is particularly important for the UltraDeep
layer, in which we are interested in very faint, distant galaxies.
The detection threshold is still effectively $5\sigma$ and we are not detecting
many fake sources.

\subsection{Artifact Rejection}
\label{sec:artifact_rejection}

We have updated the algorithm that identifies and clips transient artifacts before coaddition.
The new algorithm uses the time-series of PSF-matched warped\footnote{
  When we coadd individual visits, we resample individual CCD images to a common pixel grid
  using 3rd-order Lanczos interpolation \citep{bosch18}.  We refer to this procedure as warping
  in what follows.
}
images to identify transient artifacts, such as optical ghosts, satellite trails, and cosmic rays.
Most of the cosmic rays are identified and interpolated in the CCD processing, but some
are left unidentified and the new algorithm finds them.

The new algorithm takes both direct and PSF-matched warps as input and writes
direct coadds as output. Direct warps have been resampled to a common pixel grid.
PSF-matched warps, after being resampled to the pixel grid, have additionally been PSF-homogenized to
a Double Gaussian PSF model with a FWHM of 7.7 pixels (1.3 arcseconds), which is the seeing cut applied
in the data screening (Section \ref{sec:data_screening}) and thus all visits have better original seeing.

The new algorithm then performs the following steps.
The PSF-matched warps are stacked into a 2-sigma-clipped mean coadd which serves as a naive, artifact-free
model of the static sky.
To find artifacts, this PSF-matched sigma-clipped coadd  is subtracted from each PSF-matched warp to
produce a ``difference warp.''  Source detection is run on each difference warp to detect sources, both
positive and negative.  This step generates a set of regions (i.e., \texttt{Footprints}) for each visit,
where the pixels deviate more than 5$\sigma$ from the naive static sky model.

Some of these detections are non-astrophysical transients/artifacts to clip such
as optical ghosts and satellite trails, but other detections are not artifacts to clip.
These include astrophysical sources such as variable stars and quasars, and also
image subtraction imperfections.
These variable sources and subtraction imperfections can be separated from
the real transients using the number of epochs in which they appear. 
Variable sources and subtraction imperfections appear in most epochs because if an object is hard to
subtract cleanly in one epoch, then it is hard to subtract in most.  
This feature not only allows us to filter false positives but allows us to define ``transient'' as
a source that appears in a configurable percentage of visits. 
As a side effect, a fraction of astrophysical transients such as supernovae and
asteroids may be labeled as transient depending on observing cadence.

This temporal threshold, between ``transient`` and ``static`` is parameterized as a piecewise linear
function of the number of visits $N$.
For $N$ of 1 or 2, the threshold is  0; there is not enough information in one or two epochs to identify outliers.
For $N$ of 3 or 4, the threshold is 1, and for $N=5$, up to 2 epochs can be clipped.
For $N>5$, the threshold is $2+0.03N$ to accommodate coadds of up to hundreds of epochs.
For each artifact in each warp difference, if more than 50\% of the footprint appears in fewer visits
than this threshold, it is labeled transient.  Otherwise, it is labeled persistent and not clipped.

For more detail about the algorithm and performance compared with the clipping algorithm, see \citet{alsayyad18}.
As tested on a few tracts of PDR1 data, the new algorithm performs better in
both false positives and false negatives.

A few failure modes are known.
If an artifact persists in the coadd, it is because of one of the following reasons.

\begin{enumerate}
\item The number of epochs at its position is two or less.
  Confirm by downloading the images of the number of epochs contributing to each pixel of the coadd,
  with filenames \texttt{[patch]\_nImage.fits}.
\item In under-dithered regions, optical ghosts and chip defects overlap and appear at the same
  position in most of the exposures. The algorithm thus interprets them as part of the persistent sky,
  rather than transients.
\item The artifact is compact compared to an overlapping static source.
  These are not clipped to protect against over-clipping around stars that are susceptible to false
  positives in the image-differencing.  If the number of pixels in the footprint of a static source is
  greater than that of the artifact, the artifact is not clipped.
  This scenario occurs, for example, when a satellite trail passes through a very bright star or galaxy.
\end{enumerate}

The new artifact rejection algorithm thus works well in the Wide layer, in which dithers between
visits are large.  It is less efficient in Deep/UltraDeep layers due to smaller dithers as we
discuss in Section \ref{sec:remaining_artifacts}.

\subsection{Scattered Light in the $y$-band}
\label{sec:scattered_light_in_the_yband}

\begin{figure}
  \begin{center}
    
    \includegraphics[width=6.72cm,height=6.29cm]{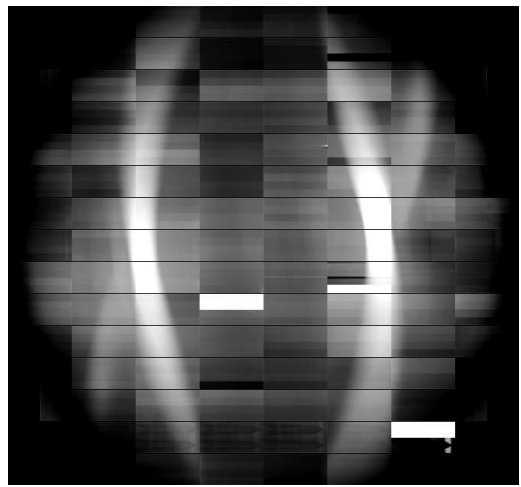}
  \end{center}
  \caption{
    Scattered light (the two vertical arcs like eyelids) found in $y$-band images.
    The image shows the whole focal plane.
  }
  \label{fig:Scattered light}
\end{figure}

Thanks to the hardware fix described in Section \ref{sec:new_filters},
scattered light from the rotator encoders is no longer seen.
However, all the $y$-band data taken before the fix suffer from it and we have developed software
to remove the scattered light.  The spatial pattern of the scattered light
changes with rotator angle in a complicated way, making analytical modeling
of it difficult.  We instead chose an empirical approach.  We obtained a sequence
of exposures by moving the rotator from $-$180 to $+$180 degrees with a step of
0.5 degrees with the shutter open and the dome closed under dark conditions.
We use these data to simulate the scattered light pattern in a given science exposure.


We first split each CCD into different read-out channels and treat each channel as
a three-dimensional array with two spatial dimensions and one periodic dimension
for the rotation angle.  We then applied the discrete wavelet transformation
(Cohen-Daubechies-Feauveau wavelet 9/7) to each of them and took level-6 approximation
coefficients along the two spatial dimensions for compression and denoising.
We achieved a compression ratio of $(2^6)^2 = 4096$, which means that
the total data volume of these exposures was reduced from 2.3TiB to 600MiB, which
is small enough to be distributed as part of the pipeline.

The scattered light subtraction procedure is as follows.
First, we compute the rotator angles at the start and end of an exposure.
Second, we load the compressed dark exposures
and interpolate them along the rotation dimension with periodic cubic splines.
We analytically integrate the cubic splines from the start angle to the end angle,
assuming that the rotator angle changes at a constant rate during an exposure.
This results in an expected illumination pattern on the CCDs that is yet to be
decompressed in the spatial dimension.  Finally, we spatially decompress
the illumination pattern, scale it to the exposure time of the image being processed,
and subtract.  This is done before the sky subtraction in the CCD processing.

\begin{figure}
  \begin{center}
    \includegraphics[width=6cm]{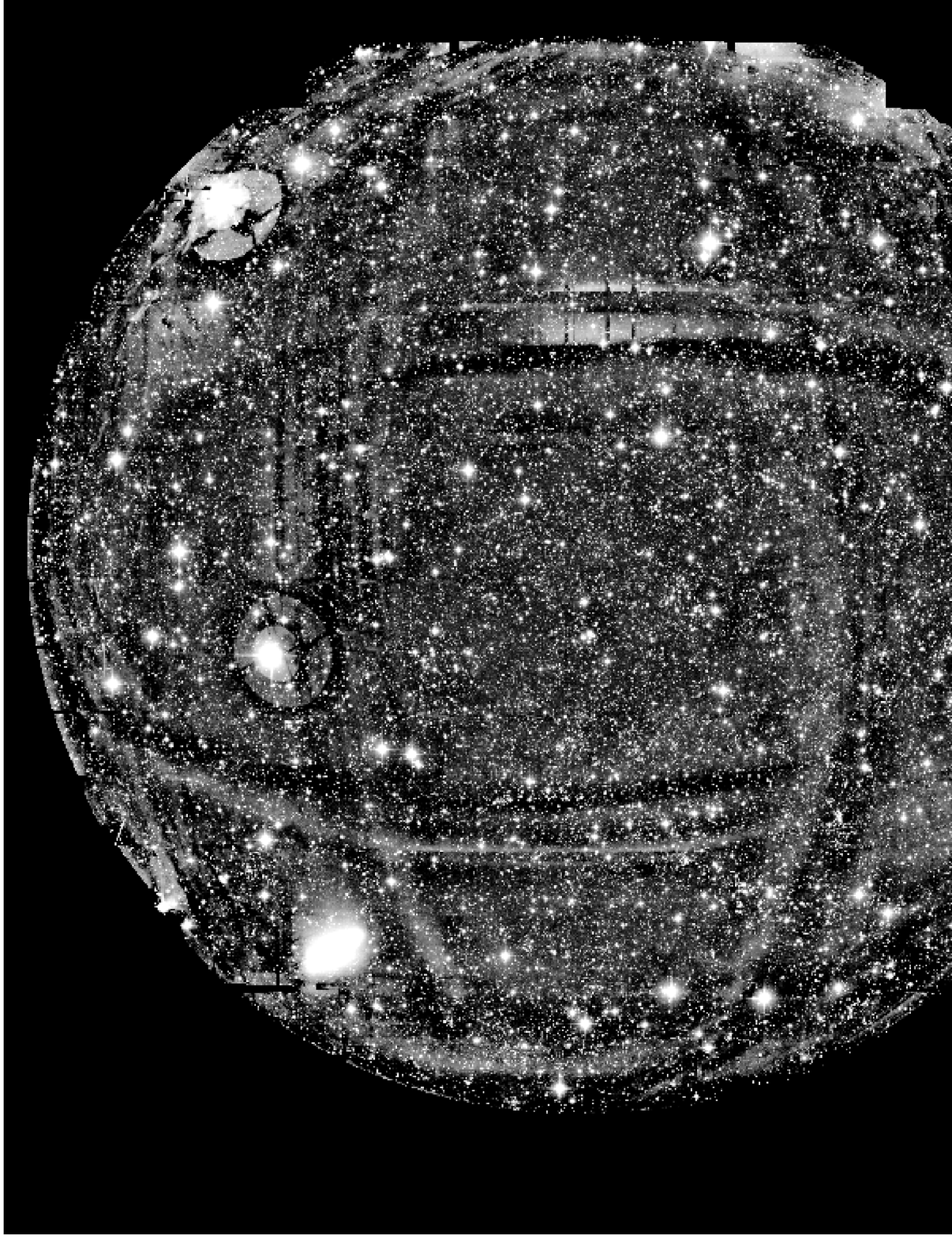}\\    
    \includegraphics[width=6cm]{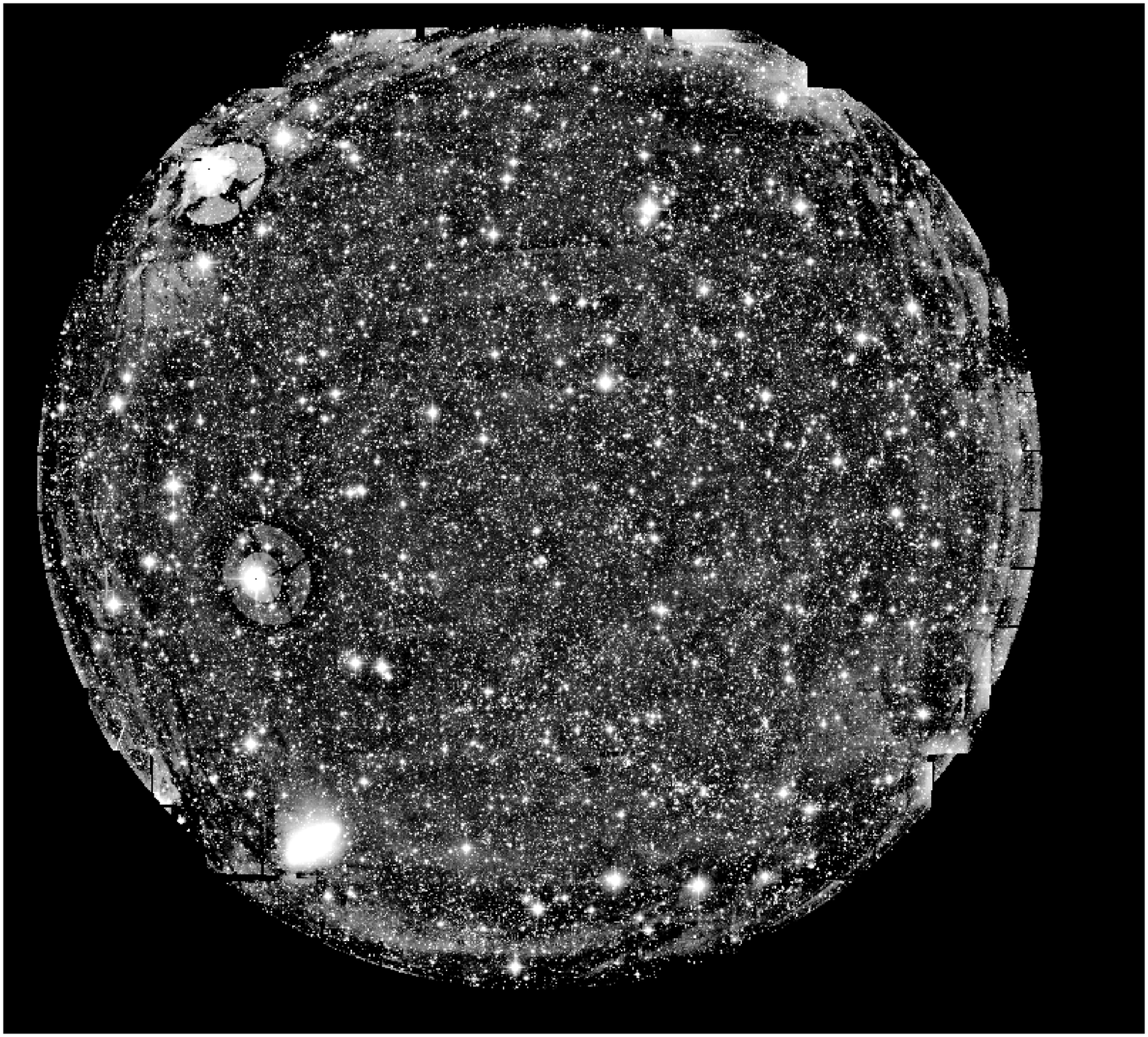}
  \end{center}
  \caption{
    {\bf Top}: $y$-band Coadd image without the scattered light subtraction.
    {\bf Bottom}: Coadd image with the scattered light subtraction.  This is only to demonstrate
    the scattered light subtraction and no careful processing has been applied (e.g., there is
    residual sky around the field edge, but that is not important).
  }
  \label{fig:Effect of the scattered light subtraction}
\end{figure}

In Figure~\ref{fig:Effect of the scattered light subtraction},
we show two $y$-band images with (bottom) and without (top) the scattered light subtraction.
This is a sample coadd image with two different position angles on the sky, differencing by 90 deg.
There is a hash-like pattern in the top image, which is due to the scattered light
remaining after the background subtraction.  The pattern is clearly gone
in the bottom image once the scattered light subtraction is performed.
The $y$-band coadd images in this release are much cleaner than those in PDR1.
We note, however, that this correction was erroneously applied to data taken {\it after}
the hardware fix; see Section \ref{sec:over_subtracted_scattered_light_in_the_yband}.

\subsection{Effective Transmission Curve}
\label{sec:effective_transmission_curve}

The image data products, both single-epoch and coadds, now contain data
structures that report an estimate of the photometric transmission as a
function of both wavelength (within a band) and position on the image.

The single-epoch transmission curves are formed by multiplying separate
spatially-constant transmission curves for the detectors, optics, and fiducial
atmosphere with bandpass filter transmission curves that vary radially over
the focal plane.  This is particularly important for the original $r$ and $i$
filters, which had strong radial dependence as discussed in Kawanomoto et al.
(2018; see also discussion in Section \ref{sec:photometry_in_i_and_i2}).
The coadd transmission curves are computed at each point on the coadd by
averaging the per-epoch transmission curves with the same weights used to
build the coadd at that point.  This process is like the PSF-model coaddition approach
described in \citet{bosch18}. This naturally reproduces the true discontinuous
spatial structure of the effective coadd transmission curve, at the expense of
a complex internal data structure.


Unfortunately, the transmission curve information is not yet utilized when
applying calibrations to the pipeline's own measurements, as this requires 
knowledge of the sub-band SEDs for objects, and the tools to infer this robustly
have not yet been developed. However, the transmission information is available in
our data products and can be exploited for scientific applications.
The easiest way for users to extract the transmission is to use tools in \code{hscPipe}
(or a compatible version of the LSST stack).  A sample script is available
at the data release site (see the FAQ page).


\subsection{PSFEx Fix}
\label{sec:psfex_fix}

\begin{figure}
  \begin{center}
    \includegraphics[width=8cm]{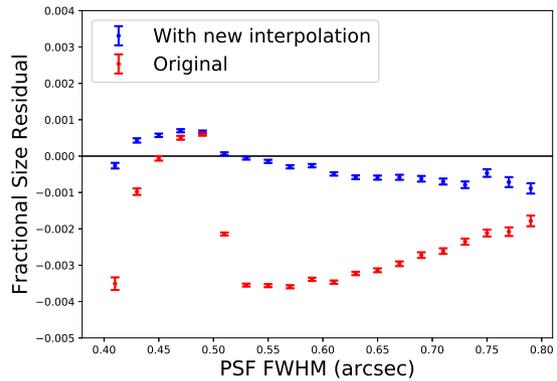}
  \end{center}
  \caption{
    Difference in the fractional size of the PSF between the observed stars and PSF model estimated using PSFex,
    as a function of seeing.
    The PSF is estimated from a number of stars in individual visits, and we then calculate the FWHM using
    an adaptive-Gaussian weighting scheme (Section 4.9.4 in \cite{bosch18}).  The red points
    use the original version of PSFex, and the width of the PSF models differs significantly from
    the widths of the actual stars.  The blue points show the results of using the modifications
    described in Section~\ref{sec:psfex_fix}.
  }
  \label{fig:psfex_fix}
\end{figure}

PSFEx (\cite{2013ascl.soft01001B}) is a widely used package for estimating an image's
point spread function (PSF).  We have repackaged it to be usable from \code{python},
and separated the choice of candidate PSF stars from the actual PSF estimation
(Section 4.3 in \cite{bosch18}).  We used a pixellated ("delta function") basis
when running PSFEx; although the individual basis functions are strongly undersampled,
fully-sampled models can still be shifted by sub-pixel offsets using sinc interpolation.
As mentioned in \citet{bosch18},
we discovered that the Lanczos kernels employed by PSFEx caused serious problems for
images with the very best seeing. 
We use a determinant radius derived from the 2nd-order moment as a measure of
the size and define a fractional size residual as

\begin{equation}
  \frac{r_{det,model} - r_{det,obs}}{r_{det,obs}},
\end{equation}

\noindent
where $r_{det,obs}$ and $r_{det,model}$ are the determinant radius of observed stars and
that of the model, respectively.  Fig.~\ref{fig:psfex_fix} shows the fractional size
residual as a function of seeing.  As the red points show, the fractional
size error increases up to 0.4\% with a sharp discontinuity at a FWHM of
around 0.5 arcsec.  We recall that the pixel scale of HSC is 0.168 arcsec per pixel.

Rather than solving the problem of determining suitable
interpolation functions, we decided to resample by interpreting the models as constant
over the sub-pixels, rather than a continuous function sampled at the pixel center.
The blue points in Fig.~\ref{fig:psfex_fix} shows the improvement by this approach.
The fractional size residual is significantly reduced and is good enough to allow us to
process all of the HSC SSP data, even those taken under the best conditions.

We can now model the PSF reasonably accurately in individual visits, but we have discovered that
the image coaddition, which comes after the individual CCD processing, introduces 
systematic errors in the PSF model.  As we discuss in detail in Section~\ref{sec:shape_measurements},
the PSF model on the coadds is larger than the observed PSF by about 0.4\%.

\subsection{Lossless Image Compression}
\label{sec:lossless_image_compression}

The total data volume of the processed HSC data has been growing rapidly
as we collect more data.  In order to save disk space,
images from the pipeline are now written using FITS tiled image compression.
The compression scheme chosen is the lossless \code{GZIP_2} algorithm
in \code{cfitsio} \citep{pence09},
which is applied to the image, mask and variance planes.
The images can be uncompressed using the \code{funpack} facility from \code{cfitsio},
or read using \code{hscPipe}.  

\subsection{Revised color terms and Restricted Color Range for Photometric Calibration}
\label{sec:colorterms}

Our photometric calibration is based on the PanStarrs (PS) DR1 data \citep{schlafly12,tonry12,magnier13,chambers16};
we apply color terms
to translate the PS magnitudes into the HSC magnitudes and perform the zero-point
calibration.  In \code{hscPipe v4} used in PDR1, we used the Bruzual-Persson-Gunn-Stryker atlas
to derive the color terms.  This atlas is is an extension of the original \citet{gunn83} atlas into both the UV and near-infrared.
For PDR2, the color terms have been updated using the newer atlas of \citet{pickles98}.
The response functions of the HSC filters used to derive the color terms have also
been updated; the old color terms were computed using the filter transmission at
the center of each filter, but we now use the filter transmission weighted by the surface area.
This operation averages the radial variation of the transmission in the $r$ and $i$-bands
and better represents the system.

There is also a change in the way we select stars for the photometric zero-point calibration.
We used to use all the stars detected in each CCD, but late-type stars tend to have a large
intrinsic color scatter primarily due to variations in metallicity.  In addition to the scatter,
there is also a systematic color offset depending on where we observe due to the stellar
population gradient across the Milky Way Galaxy.  In order to reduce such effects,
we apply color cuts to exclude late-type stars from the calibration.  To be specific,
we impose

\begin{equation}
  g-r>0\ {\rm and}\ r-i<0.5.  
\end{equation}

\noindent
The $g-r$ cut is not very important as such blue stars are quite rare.  The $r-i$ cut
eliminates stars later than K6V, which show significant color variation with metallicity.
These cuts do reduce the number of stars available for calibration.
The reduction is dependent on the sky position, but if we look at 
the COSMOS field for instance, about 40\% of the bright stars suitable for calibration
pass these color cuts and there are about 20 stars in each CCD for the zero-point calibration,
which is more than adequate for our purpose.
Comparisons with a previous internal data release do not seem to suggest a major improvement
in the zero-point uniformity, but we should in theory be more robust against metallicity variation
across the Milky Way Galaxy.

\subsection{Additional Mask Planes for Coadds}
\label{sec:additional-mask-planes-for-coadds}

As described in \citet{bosch18}, the \code{hscPipe}'s approach to PSF modeling
on coadds yields some objects that do not have a well-defined PSF, because objects
fall on a boundary such that different exposures contribute to different parts of the image.
The pipeline now includes more image-level mask planes and corresponding
catalog-level flag fields to indicate when this happens:
\begin{itemize}
    \item The \texttt{INEXACT\_PSF} image mask bit is set on any pixel for
      which the PSF is ill-defined.  The corresponding catalog flags are
      \texttt{base\_PixelFlags\_flag\_inexact\_psf}\footnote{
        The pipeline outputs are stored in the database with slightly different
        names.  The correspondence is obvious.  In this case, the flag is
        named \code{\{filter\}\_pixelflags\_flag\_inexact\_psf}, where \code{filter}
        should be the filter name such as $g$,$r$,$i$,$z$,$y$.
      }
      and \texttt{base\_PixelFlags\_flag\_inexact\_psfCenter}, where the former is
      set for any object whose above threshold detection region contains such a
      pixel, and the latter is set only when a pixel is near the center of the
      object.  Whenever \texttt{INEXACT\_PSF} is set, at least one of the
      following descriptive flags is always also set to explain why it is set.
    \item The \texttt{SENSOR\_EDGE} image mask bit is set when the pixel lies
      near the edge of at least one input image.  The corresponding catalog flags
      are \texttt{base\_PixelFlags\_flag\_sensor\_edge[Center]}.
    \item The \texttt{REJECTED} image mask bit is set on a coadd pixel when one
      or more contributing input pixels were masked during single-epoch
      processing, and could not be interpolated.  The majority of pixels with
      this mask landed on a bad amplifier or other known sensor defect.  The
      corresponding catalog flags are
      \texttt{base\_PixelFlags\_flag\_rejected[Center]}.
      \item The \texttt{CLIPPPED} image mask bit is set on a coadd pixel when one
        or more contributing input pixels were identified as belonging to
        artifacts via the image-differencing algorithm mentioned at the beginning of this section.
        The corresponding catalog flags are
        \texttt{base\_PixelFlags\_flag\_clipped[Center]}.
\end{itemize}

For many science cases, the PSF model inaccuracies reported by these flags are
actually negligible, as the PSFs of input observations are frequently quite
similar, and hence changes in which images contribute to the coadd do not
appreciably affect the PSF.  We encourage science users to test their
analysis both with and without filtering on these flags to determine whether
this effect is important for those cases.



\section{Data}
\label{sec:data}

We mostly followed the same data processing procedure as in PDR1, but we briefly describe
how we screened the data and processed them with an emphasis on the differences from PDR1.

\subsection{Data Screening}
\label{sec:data_screening}

We applied mostly the same conditions to screen the raw data for the full
processing as in the last release (see Section 3.2 of \cite{aihara18b}).
The main conditions were (1) sky brightness $\leq 45000$ ADUs,
(2) seeing $\le 1.3$ arcsec, and (3) sky transparency $\geq 0.3$.
The $y$ and NB921 filters are affected by sky fringing, which needs to be subtracted off.
To generate fringe patterns in these filters, we used a slightly relaxed condition:
seeing below $1.5$ arcsec.
Our new sky subtraction algorithm (Section \ref{sec:global_sky_subtraction}) uses
data for the sky frame selected using the same relaxed condition.
For both fringe and sky frames, we generated
a master frame for each filter for each observing run. If the number of visits available
in a run is insufficient ($<50$), we combined data from a few nearby observing runs.
We performed careful visual inspections of the coadd images in
addition to the automated screening of individual visits and reprocessed
several tracts with problematic visits removed.  For example, we removed 5 tracts in the $r$-band
that were accidentally traversed by laser light for adaptive optics from another telescope.

\subsection{Data Processing}
\label{sec:data_processing}

The processing flow remained largely the same as in the last data release, but
some small changes were made to incorporate the new features.

We first generated calibration images for the bias and dark subtraction, flat-fielding, 
fringe subtraction, and global sky subtraction.  Each raw CCD image was
processed by applying these calibration data.  The removal of the $y$-band scattered
light described in Section~\ref{sec:scattered_light_in_the_yband} was also performed here (prior to the sky subtraction).
The astrometric and photometric calibrations were
carried out against the Pan-STARRS1 DR1 catalog.  In this CCD processing,
the same configuration parameters were used in all the filters, except for NB387,
which is the least sensitive filter and had many fewer stars available for calibration.
We lowered the star selection threshold to include fainter stars so that the processing
did not fail.
The next step, global sky subtraction (Section~\ref{sec:global_sky_subtraction}), was
a big change from PDR1.  This stage was implemented as a separate process and produced
sky images that were subtracted from the calibrated CCD images.

After the individual CCD calibrations, we performed a multi-visit calibration of
astrometry and photometry to refine the CCD calibrations.  Then, the CCD images
were warped onto common coordinate grids and combined to generate deep coadds.
This was done for each band separately, but we combined the $i$ and $i2$,
and $r$ and $r2$-bands as mentioned earlier.  Objects were detected on the coadds,
and detections from multiple filters were merged to a single detection catalog.
The final stage, multi-band measurements, performed object deblending and
various photometric measurements.  The resultant multi-band catalog is the one
most useful for science.  PSF-matched aperture photometry was mistakenly excluded
from this last step and were ran as an afterburner process.  It has been merged with
the other measurements at the database.


\subsection{Image and Catalog Data}
\label{sec:image_and_catalog_data}

The pipeline generates calibrated CCD images, images warped to patches (warps), and coadds as well as the associated catalog files.
They are all available from our website.  Once again, users should be
aware that the lossless compression has been applied to the image files
(see Section \ref{sec:lossless_image_compression}).  The directory structure of
the pipeline outputs is similar to PDR1, but the exact locations of some of
the files are different.  An important change for the users to notice is that the final coadds are now under
the \code{deepcoadd-results/} directory (PDR1 had them under \code{deepcoadd/}).
There are several new files, which we describe at the data release site.


Galaxy shape measurements based on the \citet{hirata03} algorithm were withheld from PDR1.
As some of the flat files contained the HSM shape measurements, these files were also withheld.
It was originally meant to be a short-term solution, but PDR1 ended up withholding these files
for about 2 years. A major side effect was that other useful information in those files were
inaccessible from users (e.g.,  single-epoch source catalogs that were not loaded to the database).
In this release, we again choose to withhold the shape measurements as well as the deblended images,
but we make all flat files available, where we exclude only the shapes and deblended images from the files.

We remind the reader again that the shape measurements from PDR1 are now available
as part of this PDR2 (see Section \ref{sec:calibrated_shape_measurements}).  All
the flat files withheld from PDR1 are available, too.  Some of the most useful files may
include {\tt meas} files, which include deblended images, and {\tt CALSRC} catalogs,
which are single-epoch source catalogs.  See the PDR1 site for the list of flat files.

\subsection{Value-added Products}
\label{sec:value_added_products}

In addition to the main data set described above, we include a few value-added data
products in this release.

\begin{itemize}
\item {\bf COSMOS Wide-depth stacks:}
  There are many visits in UD-COSMOS observed under a wide range of seeing conditions.
  We have stacked a subsample of the UD-COSMOS visits to the nominal exposure times of the Wide survey
  for 3 different sets of seeing conditions.  In PDR1, we made coadds with 0.5, 0.7 and 1.0 arcsec
  seeing, but in PDR2, we instead coadd UD-COSMOS visits with target seeing at
  25\%, 50\%, and 75\% of the seeing distributions in each filter in the Wide layer.
  The seeing for each filter for each stack is summarized in Table \ref{tab:wide_depth_seeing}.
  The target seeing is chosen based on the S17A internal data, and the seeing in Table  \ref{tab:wide_depth_seeing} is not
  fully consistent with the median and quartile seeing summarized in Table \ref{tab:exptime}.
  Nontheless, these Wide-depth stacks will be useful for characterizing the data quality variation in the Wide layer.
  The multiband photometry for each stack is of course available.
  
\item {\bf Public spectroscopic redshifts:}
  We have updated the list of public spectroscopic redshifts from the literature.  The list includes
  redshifts from zCOSMOS DR3 \citep{lilly09}, UDSz \citep{bradshaw13,mclure13},
  3D-HST \citep{skelton14,momcheva16}, FMOS-COSMOS \citep{silverman15,kashino19},
  VVDS \citep{lefevre13}, VIPERS PDR1 \citep{garilli14}, SDSS DR12 \citep{alam15}, the SDSS IV QSO catalog \citep{paris18},
  GAMA DR2 \citep{liske15}, WiggleZ DR1 \citep{drinkwater10},
  DEEP2 DR4 \citep{davis03,newman13}, DEEP3 \citep{cooper11,cooper12}, and
  PRIMUS DR1 \citep{coil11,cool13}. 
  As one-to-one correspondence between the spectroscopic objects and photometric objects is not always
  obvious, we match objects within 1 arcsec and all matched objects are stored in the database.
  In most cases, the  most likely match will be the object with the smallest matching distance.
  There is also
  a homogenized spectroscopic confidence flag for each object to make it easy for users to make a clean
  redshift catalog (recall that each spectroscopic survey has its own flagging scheme).  See the online documentation
  for the definition.  We emphasize that users should acknowledge the original data source(s) when using this table.
  
\item {\bf Random points:}
  We draw random points with a density of 100 points per square arcmin for each coadd image for each filter.
  These random points can be used for e.g., clustering analysis, identifying problematic areas, computing
  the survey area and the fraction of masked areas, etc.  The random points are available in the database
  and the data release site describes how to use this table.
  Note that the random points are affected by the issues with masks around bright stars (Section \ref{sec:bright_star_masks}).
  We plan to update the random point catalog together with the revised masks in a future release.
\end{itemize}

We have also computed photometric redshifts for a large number of objects. They are not included
in the current release but will be released in a future incremental release.  Other data products
may also be released and we will make announcements on the data release website.  Registered users
will be notified.

\begin{table}[htbp]
  \begin{center}
    \begin{tabular}{cccccc}
      \hline
      Stack     & $g$  & $r$  & $i$  & $z$  & $y$\\
      \hline
      Best      & 0.63 & 0.61 & 0.52 & 0.70 & 0.59\\
      Median    & 0.74 & 0.79 & 0.57 & 0.75 & 0.73\\
      Worst     & 0.87 & 0.89 & 0.67 & 0.83 & 0.94\\
      \hline 
    \end{tabular}
  \end{center}
  \caption{
    Seeing in arcsec for each filter and for each of the COSMOS Wide-depth stack.
  }
  \label{tab:wide_depth_seeing}
\end{table}

\section{Data Quality and Known Issues}
\label{sec:data_quality_and_known_issues}

We now demonstrate the quality of the data in this release.  There have been a number of
pipeline changes since PDR1 as described above and we believe that the overall quality is significantly improved.
We have performed an extensive set of quality assurance tests in our validation campaign.
In what follows, we present some of the most important tests with one or two key figures for each test.
A full set of tests and figures can be found at the data release site.


\subsection{Photometry: Internal Consistency}
\label{sec:internal_consistency}

\begin{figure}
  \begin{center}
  \includegraphics[width=8cm]{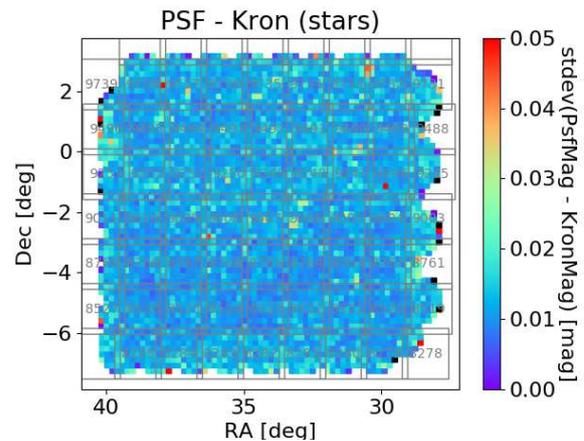}
  \end{center}
  \caption{
    Scatter between Kron and PSF magnitudes in the $g$-band for bright ($i<21.5$) stars in the Wide XMM-LSS field.
    The scatter is measured separately in each patch.  The large squares show the tract borders.
  }
  \label{fig:kronpsf_diff}
\end{figure}

We first present the photometric quality of our data.  We begin with internal consistency checks.
In Fig.~\ref{fig:kronpsf_diff}, we compare the difference between Kron magnitudes and PSF magnitudes
for bright ($i<21.5$) stars in the Wide XMM-LSS field in the $g$-band. The PSF photometry is
based on the model PSF constructed by coadding PSFs from individual visits \citep{bosch18}, while the Kron photometry
is a moment-based adaptive-aperture measurement on the coadd.  Thus, the consistency between them is a good measure
of the internal consistency.  As shown in the figure, we achieve $\sigma\sim0.01$ mag across
the field.  This level of scatter is typical of most fields in most filters, except for the $y$-band, which shows
a larger scatter ($\sigma\sim0.015$), due at least in part to shallower depths.
Overall, this test demonstrates good internal accuracy of our photometry. We also perform the same
analysis comparing the CModel and PSF photometry.  CModel asymptotically approaches PSF for
point sources and we indeed achieve an excellent performance of $\sigma\lesssim0.002$ mag (plot not shown).
These trends are largely filter-independent, although they do depend on the seeing size as expected.
All the plots for each field and filter are available online.

\subsection{Photometry: External Consistency}
\label{sec:external_consistency}

\begin{figure*}
  \begin{center}    
    \includegraphics[width=8.5cm]{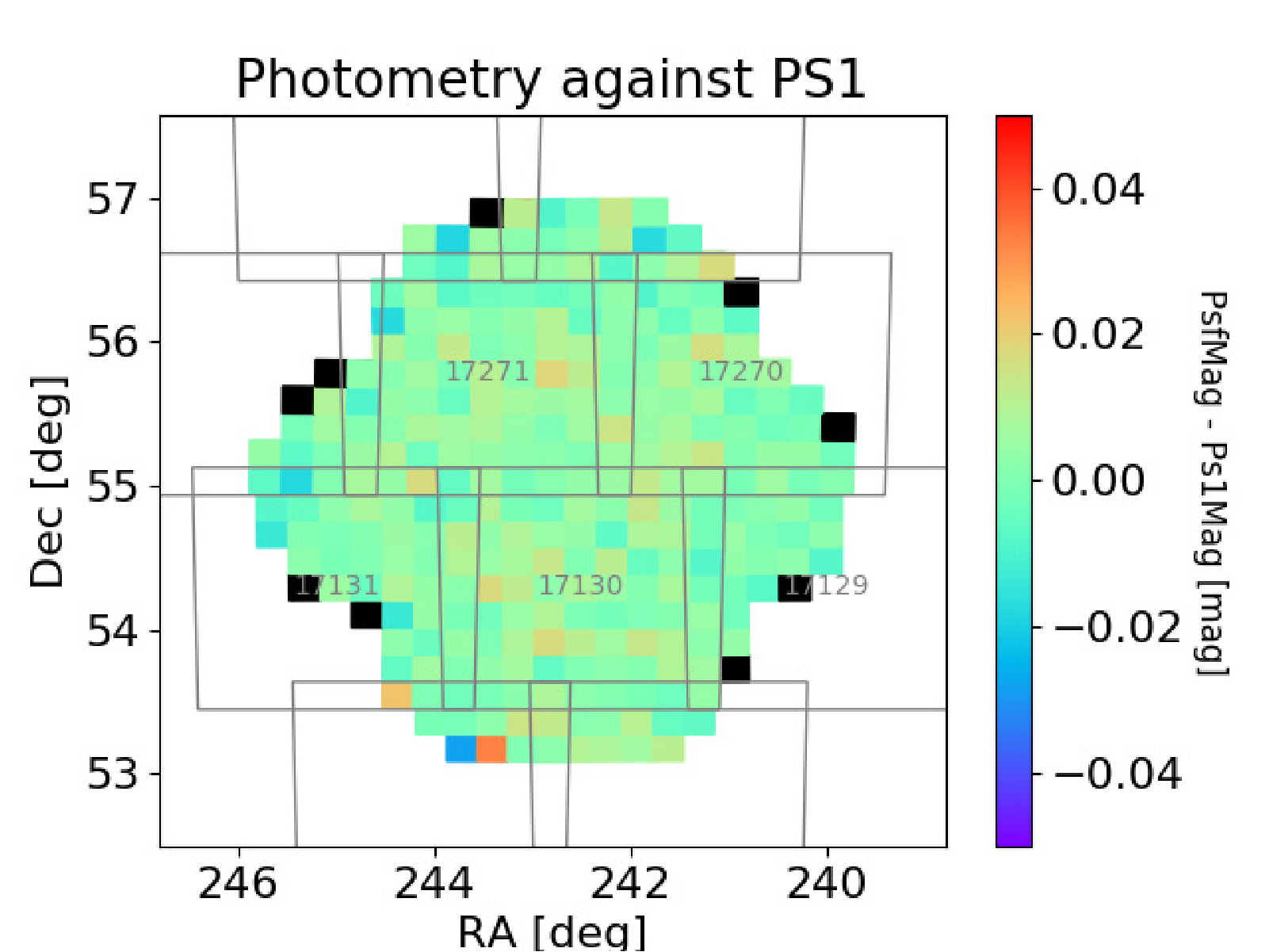}
    \includegraphics[width=8cm]{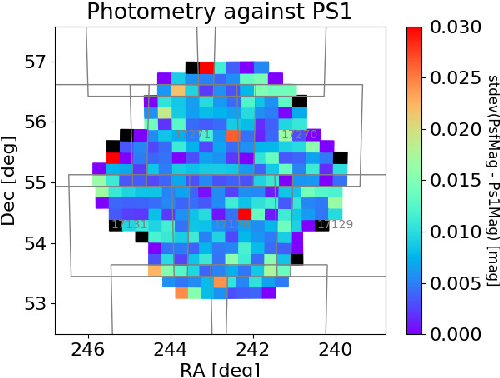}
  \end{center}
  \caption{
    {\bf Left:} $r$-band PSF magnitude difference between HSC and PS1 for point sources in the Deep ELAIS-N1 field.
    The difference is measured in each patch separately.  The large squares show the tract borders.
    {\bf Right:}
    Same as left panel, but the color scale shows the scatter.
  }
  \label{fig:psfmagdiff_ps1}
\end{figure*}

Next, we make comparisons between HSC and external data sets.  Fig.~\ref{fig:psfmagdiff_ps1}
compares the $r$-band photometry between HSC and PS1 for point sources with $r<20$.
This magnitude cut is brighter than that applied in the previous figure because we now compare
with the shallower PS1 data.
We apply a color term to translate the PS1 system into the HSC system for
a fair comparison.  As we use the PS1 photometry to calibrate the HSC zero-points, the comparison
here is not entirely external but it is still useful.

The left figure shows that the difference between HSC and PS1 is close to zero, as expected.
Most filters and fields do not show a significant difference, but there is a small offset in
the $i$-band in the Deep+UltraDeep fields and also in some small regions in the Wide fields.
This is due to a calibration issue discussed in Section \ref{sec:photometry_in_i_and_i2}.
The right figure shows a scatter of the magnitude difference.  The scatter is at a level of
$\sim0.01$, indicating a good photometric consistency.  A similar level of consistency is seen
in most of the other filters and in other fields.  The $y$-band generally shows a slightly larger scatter ($\sim0.02$ mag)
possibly due to the varying water vapor absorption in the atmosphere.  The comparison is much worse
in NB387 ($\sim0.2$ mag), but this is due to the intrinsic color scatter of stars;
the color terms to extrapolate from PS1 photometry to NB387 are sensitive to stellar metallicity variations,
which the PS1 photometry cannot fully capture.
It should be noted that there are small regions where the scatter is significantly larger in the broad-bands
(see Section \ref{sec:inconsistent_fluxes}).

\begin{figure*}
  \begin{center}
  \includegraphics[width=15cm]{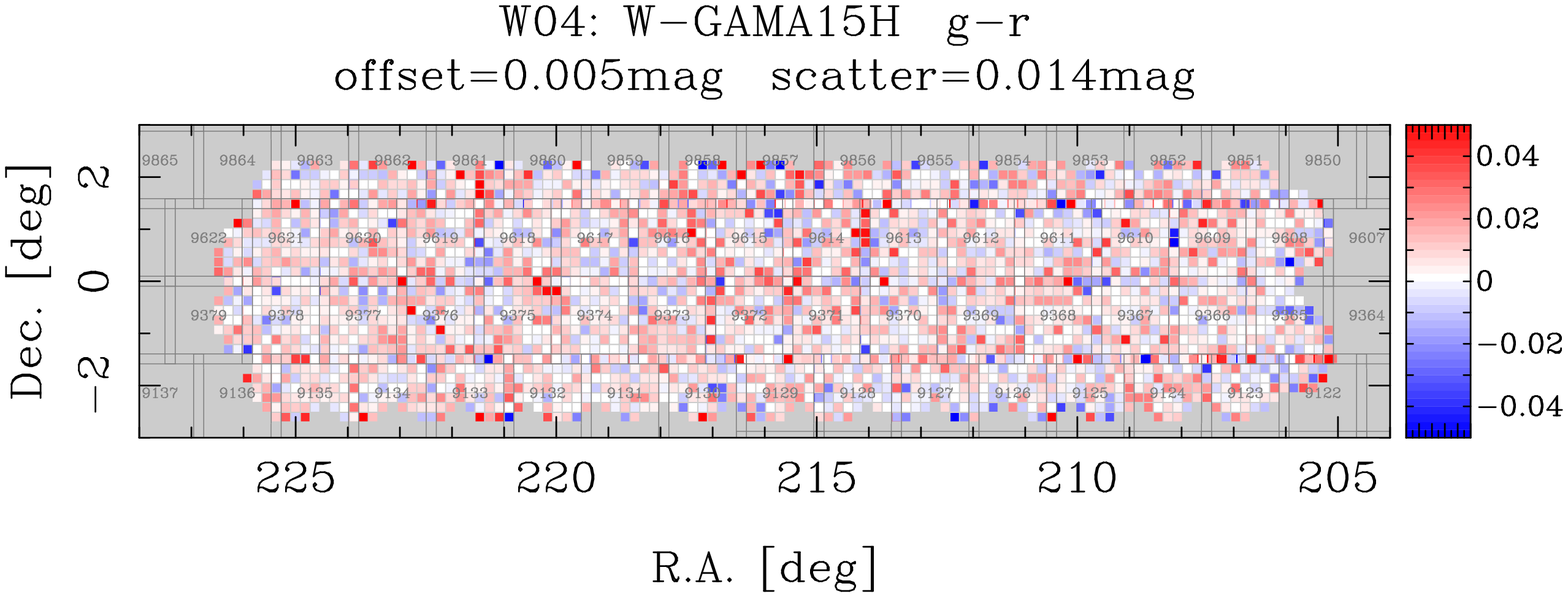}
  \end{center}
  \caption{
    $g-r$ color difference between HSC and PS1 in the Wide-GAMA15H field.
  }
  \label{fig:color_comp}
\end{figure*}

Fig.~\ref{fig:color_comp} shows color differences between HSC and PS1.  Here we choose the Wide-GAMA15H field
in the $g-r$ color as an example. The other colors and fields can be found online.  As expected from the magnitude differences
discussed above, we observe no significant color offset and no spatial structure in the $g-r$ color.
The color scatter is small, $\sim1.5\%$, which is reassuring.

\begin{figure*}
  \begin{center}
    \includegraphics[width=12cm]{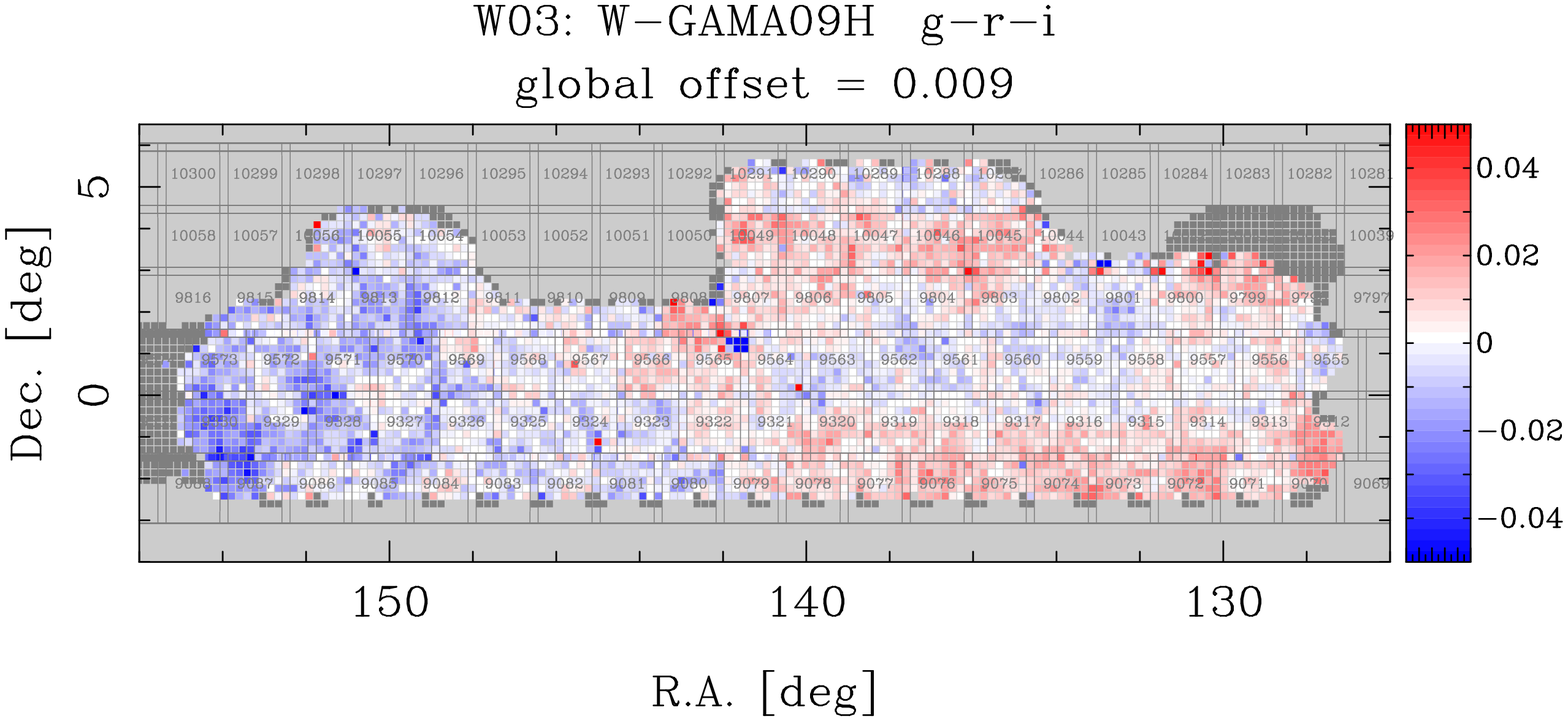}\\\vspace{1cm}
    \includegraphics[width=12cm]{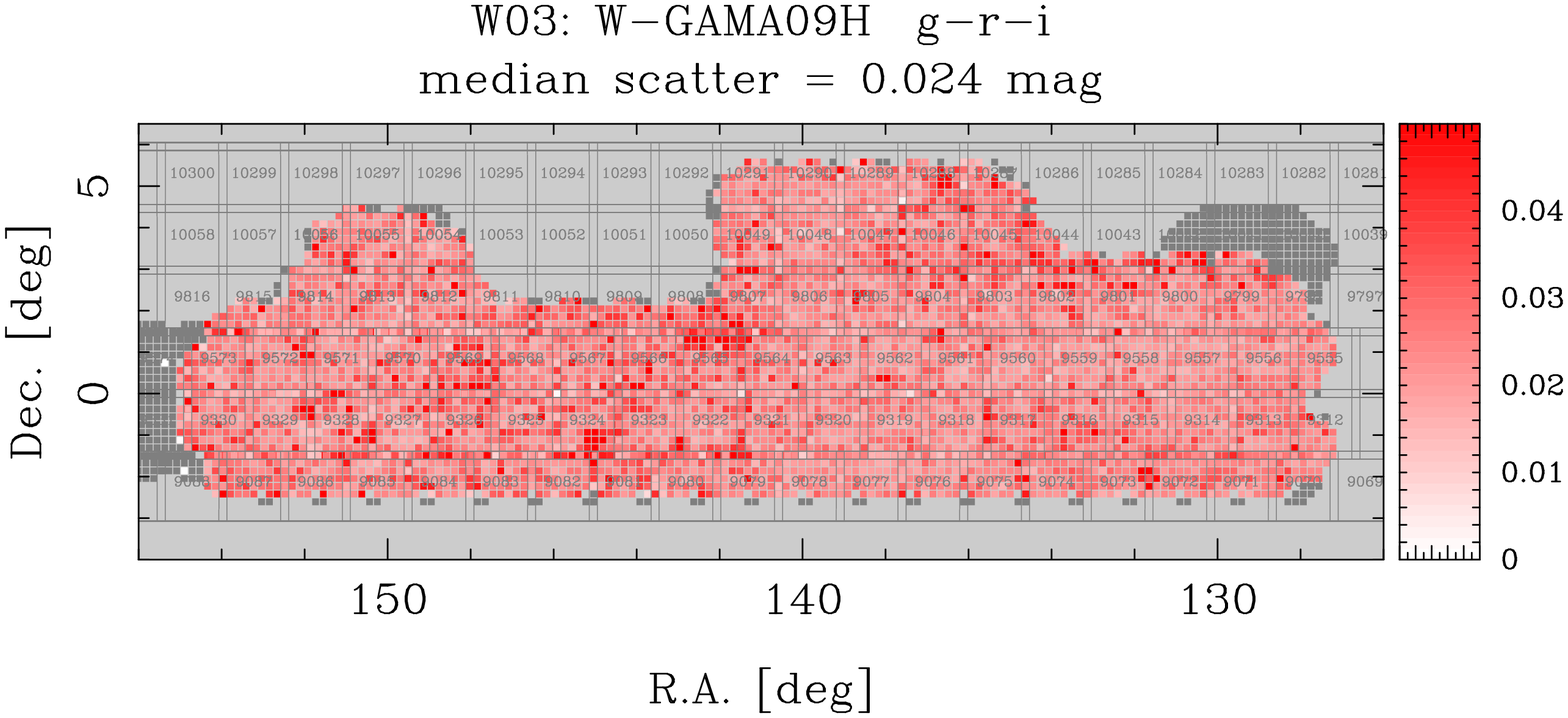}
  \end{center}
  \caption{
    {\bf Top:} Color offset of the stellar sequence with respect to the expected sequence
    from the \citet{pickles98} stellar library.  The color bars on the right shows the level
    of offset.  The dark gray regions are either the area where we do not have a sufficient
    number of stars (mostly field edges) or the area that is not covered in all the required
    filters (e.g., $gri$ in this case).  In order to enhance the spatial (non-)uniformity,
    we subtract the median offset over the entire field.
    {\bf Bottom:} As in the top panel but for the color scatter of the stellar sequence.
    The small squares represent patches and the large ones represent tracts.
    The median scatter over the field is indicated in the plot.
  }
  \label{fig:stellar_sequence}
\end{figure*}

Another useful external check is to compare the observed location of the stellar sequence on
a two-color diagram with the location expected from a stellar spectral library.  We use
the \citet{pickles98} library as above and compute the synthetic magnitudes for each filter
using the HSC total system response functions.  We fit 2nd order polynomials to the linear
part of the synthetic stellar sequence, avoiding late-type stars.  We then estimate the offset
between the observed stars and the fitted curve.  This is done for each patch separately.
If our calibration is good, we expect that
the offset is small (but not necessarily be zero because the stellar library likely has small but non-zero systematics)
and uniform over the entire survey area.  The top panel of
Fig.~\ref{fig:stellar_sequence} shows a sample plot for the GAMA09H field using the $gri$ photometry.
The offset is mostly within $\sim0.02$ mag, indicating a good calibration, although there is
weak spatial structure.
The structure at R.A.=130-140 deg is likely due to the $i$ vs $i_2$ issue discussed in Section \ref{sec:photometry_in_i_and_i2}.
We have applied corrections for the Galactic extinction, but not all the stars are completely
behind the dust curtain and the correction may introduce a spatial structure.
This may actually explain the feature around R.A.=150 deg. as we do not observe a significant
difference between HSC and PS1 magnitudes there (we do not correct for the extinction in
the HSC vs. PS1 comparisons).
We further measure the scatter around the observed stellar sequence
as it is another good indicator of the photometric accuracy.  The observed scatter of the stellar
sequence plotted in the bottom panel is $\sim0.02$ mag and is fairly uniform across the field.
Note that the scatter is due to the three filters, suggesting that the scatter in calibration errors
is roughly $0.02/\sqrt{3}$ per filter.
Overall, these tests suggests that our photometric calibration is accurate to about 1\%, which should be
sufficient to enable a wide variety of scientific explorations of the data.
The current accuracy is reasonably good, but we expect to achieve better accuracy in the future with
the effective transmission curve discussed in Section \ref{sec:effective_transmission_curve}.

\subsection{Astrometry}
\label{sec:astrometry}

\begin{figure*}
  \begin{center}
    \includegraphics[width=16cm]{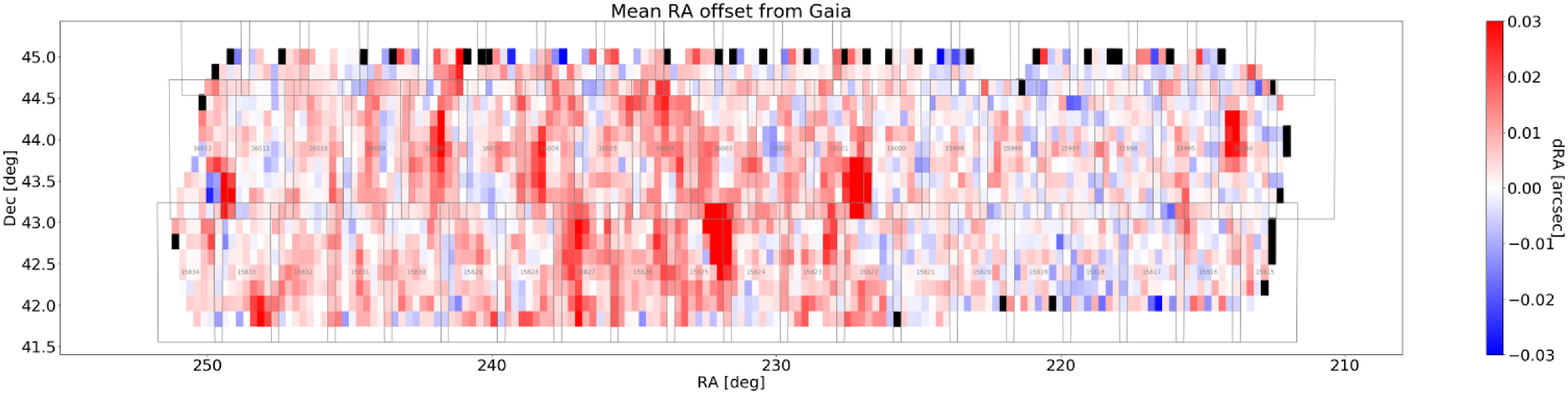}\\\vspace{0.5cm}
    \includegraphics[width=16cm]{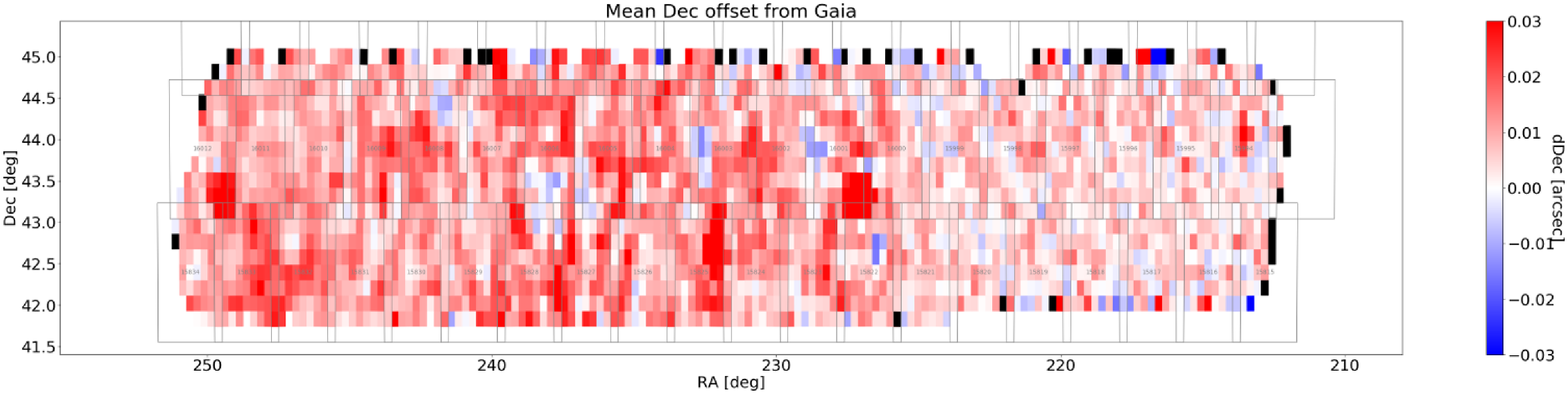}
  \end{center}
  \caption{
    Astrometric differences between HSC and Gaia in the $i$-band.  Mean offset per patch in
    Right Ascension and Declination are plotted in the top and bottom panels, respectively.
    The mean here is computed as $3\sigma$ clipped mean, and is thus robust against outliers.
  }
  \label{fig:astrometry}
\end{figure*}

The astrometric catalog from Gaia is an obvious choice of external source to
evaluate the astrometric accuracy of the HSC data.  As in the previous section,
we use bright point sources to estimate the astrometric errors relative to Gaia DR1
\citep{gaia16a,gaia16b}\footnote{
  At the time when we processed the PDR2 data, the Gaia DR2 was not yet available.
  We will use the DR2 (or newer) in our future release.
}.
As an example,
we show the Hectomap field in Fig.~\ref{fig:astrometry}, where we show the mean offset in
position in each patch.  There seems to be a large-scale
trend that the offset becomes larger at larger Right Ascension.  Such large-scale features are
also observed in other fields.  In addition, there are a few small regions, typically
a chunk of several patches, where the offset
is larger than the others.  Such small-scale features are seen in the other fields as well.
We have not yet understood where these features come from, but they might possibly due
to the PS1 catalog because the spatial pattern does not follow the tract borders
(we apply a tract-wide astrometric calibration in the processing).
They could also be due to proper motions, which we have ignored in our astrometric
calibration process.
In any case, even the large offsets are below 0.1 arcsec and most science
cases are unlikely to be significantly affected by these astrometric errors,
but users who require high precision object positions should be warned.

\subsection{PSF Model}
\label{sec:shape_measurements}

\begin{figure}
  \begin{center}
    \includegraphics[width=8cm]{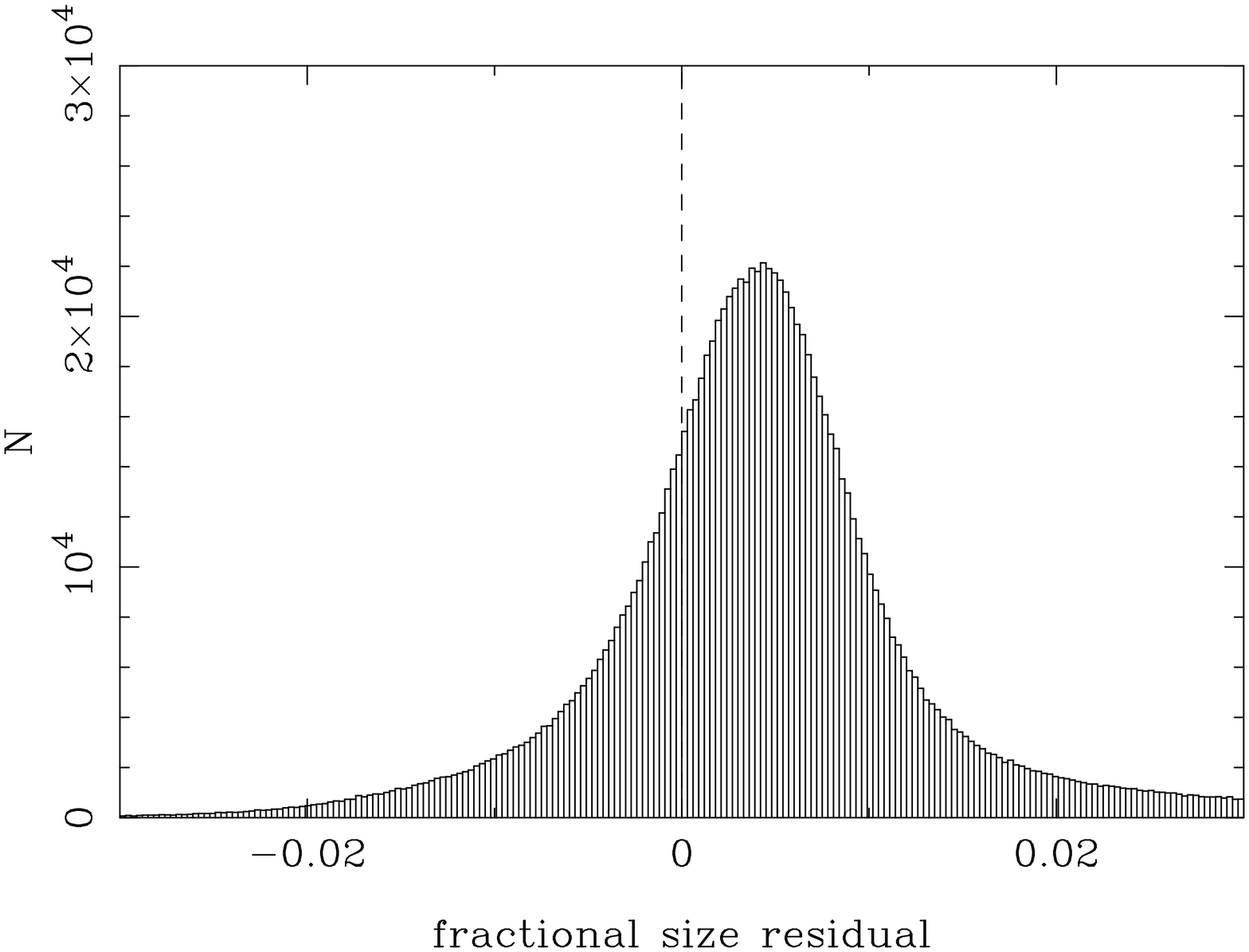}
  \end{center}
  \caption{
    Distribution of the fractional size residual in the $i$-band for bright stars ($i<22$) in the Wide-VVDS field.
    The vertical dashed line shows the zero residual.  The peak is positive, meaning that the width of
    the modeled PSF is biased slightly large.
  }
  \label{fig:size_residual}
\end{figure}

A key ingredient for precise shape measurements is a good PSF model.
As described in \citet{bosch18}, our PSF model at a given position on a coadd is
constructed by coadding the PSF model from individual visits.  One easy test of the
PSF model accuracy is to compare the size of the observed PSF in the coadds with that of the model
PSF at the same position.
Fig.~\ref{fig:size_residual} makes this comparison.  We use the fractional size residual defined
in Eq. 2 and use only bright ($i<22$) stars.
The residual is small, but not
zero.  The residual is about $4\times10^{-3}$, meaning that the model PSF is larger
than the observed PSF.  This is unlikely to affect the object detection and photometry
at a significant level, but it does not pass our stringent requirement for cosmic shear analysis
(see discussion in \cite{mandelbaum18}).
Investigations are underway to fully understand this residual.
We have found that at least part of it is from
image warping: we warp individual CCD images with the third-order Lanczos kernel
when we generate coadds, but the size residual decreases if we increase the order to fifth order.
However, this does not seem to fully solve the problem and further work is needed here.
When we release the shape measurements from PDR2, we hope to release an updated version of
the PSF model that passes the cosmic shear requirement.
We note that galaxy shapes may be better measured in individual CCD
images than on coadds as it avoids the need for pixel resampling
during warping. However, the current residuals in the PSF models
obtained on the coadds are not sufficiently large to motivate the
substantial effort of developing, validating and using such an
approach. Preliminary evidence suggests that modest adjustments to our
current approach of measurements on the coadds would likely be
successful in reducing the PSF model residuals below our science
requirements for current as well as future releases (Armstrong et al. 
in prep.).

\subsection{Survey Depth}
\label{sec:survey_depth}

\begin{figure}
  \begin{center}
    \includegraphics[width=8cm]{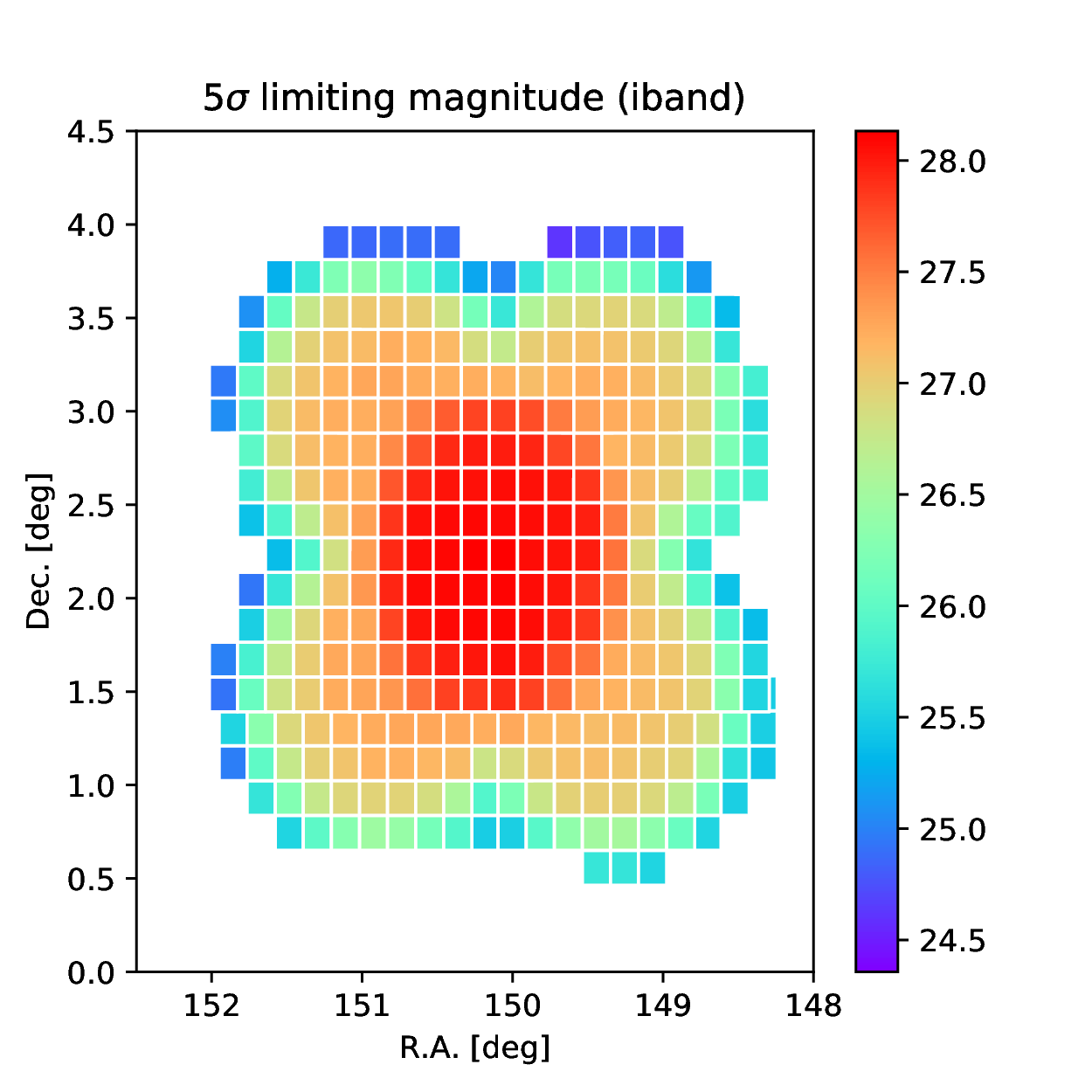}
  \end{center}
  \caption{
    $5\sigma$ limiting depth in the $i$-band for point sources in D/UD-COSMOS evaluated separately
    in each patch.  
  }
  \label{fig:depth}
\end{figure}

\begin{figure*}
  \begin{center}
    \includegraphics[width=16cm]{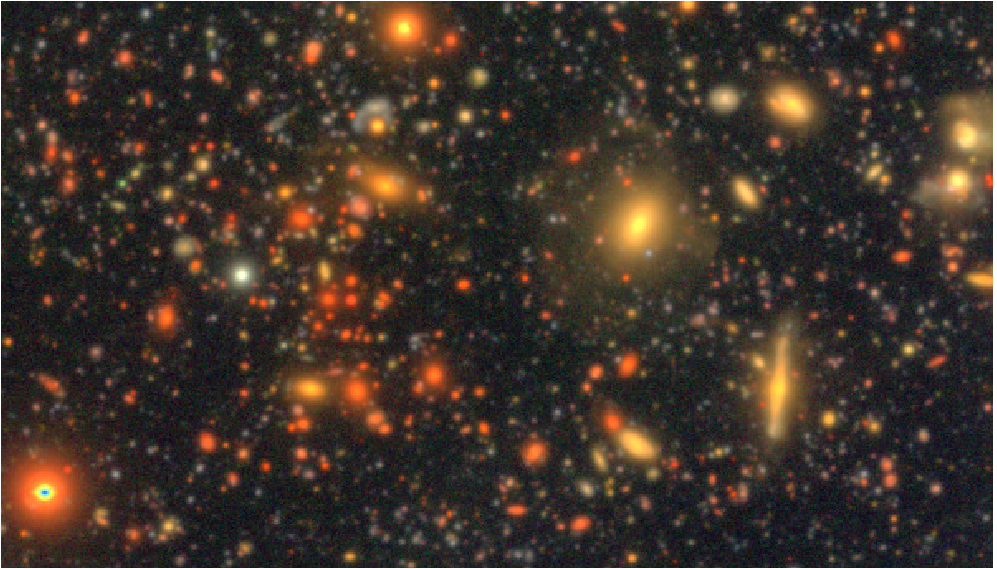}
  \end{center}
  \caption{
    $gri$ color-composite of a small chunk ($3'.5\times2'.0$) of the COSMOS field centered at
    R.A.=$\rm10^h00^m20^s.0$, Dec.=$\rm+02^\circ11'55''.0$.  North is up.  This image is
    colored following the algorithm of \citet{lupton04}.
  }
  \label{fig:cosmos}
\end{figure*}

Another useful quantity to characterize imaging data is the depth.  There are multiple ways to
define the depth, but we adopt a simple definition here; $5\sigma$ limiting magnitude
for point sources.  We evaluate the limiting magnitudes using stars with $S/N$ between
4.9 and 5.1 for each patch.  We use the $S/N$ quoted by the pipeline, which is likely a slightly
optimistic estimate due to the ignored pixel-to-pixel covariance as discussed earlier.
Fig.~\ref{fig:depth} shows a map of the $i$-band depth in
the Deep/UltraDeep COSMOS field.  The Deep and UltraDeep data were jointly processed as mentioned
earlier, giving rise to spatial structure in the figure.  In the central pointing,
we reach $i\sim28$; this is the deepest optical image of the field in existence.
Fig.~\ref{fig:cosmos} is a nice illustration of the depth we reach; there are so many
objects in this small cutout that there is almost no empty space between the objects.
The depths in the other fields are fairly uniform, although there is also some spatial
structure due to combination of dithering pattern and seeing variations.  See the QA page
of the data release site for more plots.

\subsection{Known Issues}
\label{sec:known_issues}

Thanks to the updated processing pipeline, the overall data quality is improved since PDR1.
However, there are some persistent problems and also new problems.  This section
summarizes the problems known to date.  We will not repeat the problems that persist from PDR1 here;
bright galaxy shredding and underestimated flux uncertainties
in convolved measurements are discussed in Sections 5.8.3 and 5.8.11 of \citet{aihara18a}, respectively.
Optical ghosts due to bright stars (Section 5.8.8 of \citet{aihara18a}) are significantly reduced, but
they are not completely gone and will be briefly discussed here.
The issue of deblending failures in crowded areas (Section 5.8.10 of \cite{aihara18a}) also persists,
but there is a pipeline change related to it and we will discuss it here as well.
The list of known issues will evolve with time; we will keep the list up to date at the data release site.

\subsubsection{Remaining Artifacts on Coadds}
\label{sec:remaining_artifacts}

\begin{figure}
  \includegraphics[width=0.49\textwidth]{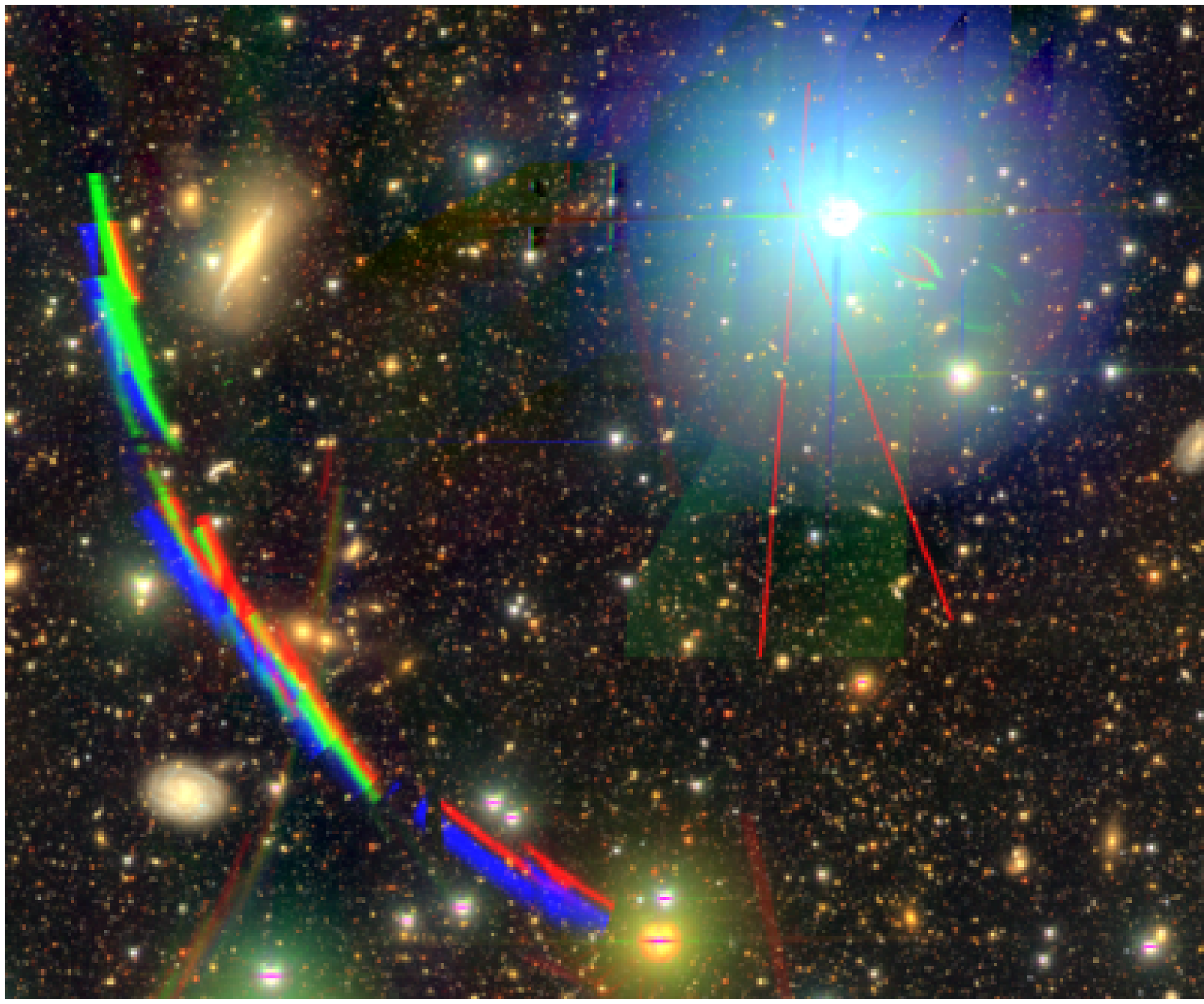}
  \caption{
    Remaining optical ghosts and satellite trails in UD-COSMOS.  The image is in $gri$ and is approximately $23'\times19'$.
  }
  \label{fig:ghosts}
\end{figure}

The artifact rejection algorithm mentioned in Section \ref{sec:artifact_rejection} is
very effective, particularly in the Wide survey in which the dither is large
(approximately 1/3 of field of view).  However, it is less
effective in UltraDeep: the dithers are smaller (several arcmin) and optical ghosts
stay roughly at the same position on the sky, making the artifact rejection based
on the image differencing difficult.  Fig.~\ref{fig:ghosts} shows an example case in UD-COSMOS.
There are a few satellite trails remaining there as well, but they are due to the enhanced background from
bright star nearby (this is one of the failure modes discussed in Section \ref{sec:artifact_rejection}).
Work is in progress to predict the locations of optical ghosts using
the optical model of the instrument as well as to identify satellite trails using
the Hough transform.  We expect that artifacts will be further reduced in future data releases.

\subsubsection{Bright Star Masks}
\label{sec:bright_star_masks}

\begin{figure*}
  \begin{center}
    \includegraphics[width=8cm]{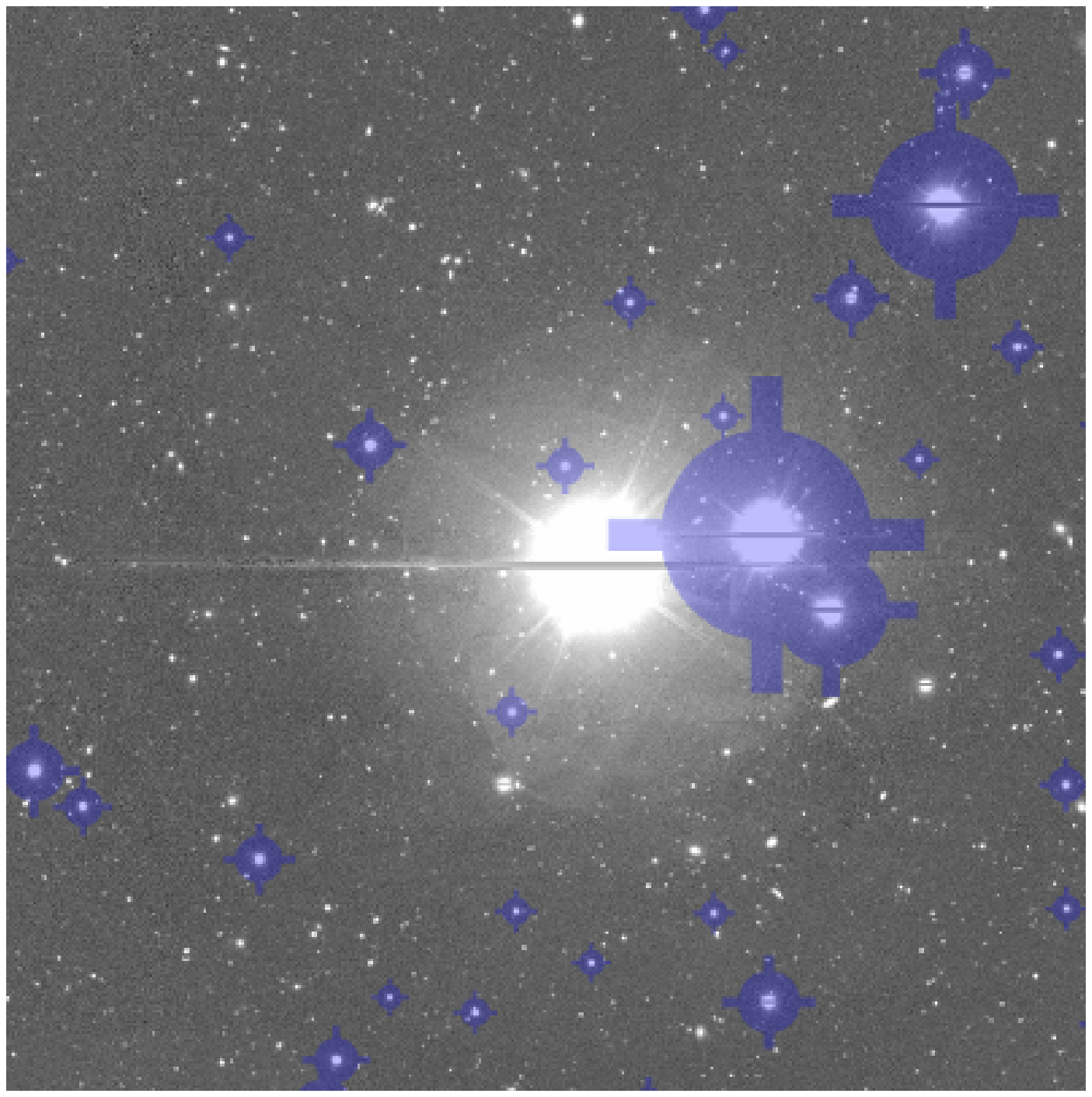}
    \includegraphics[width=8cm]{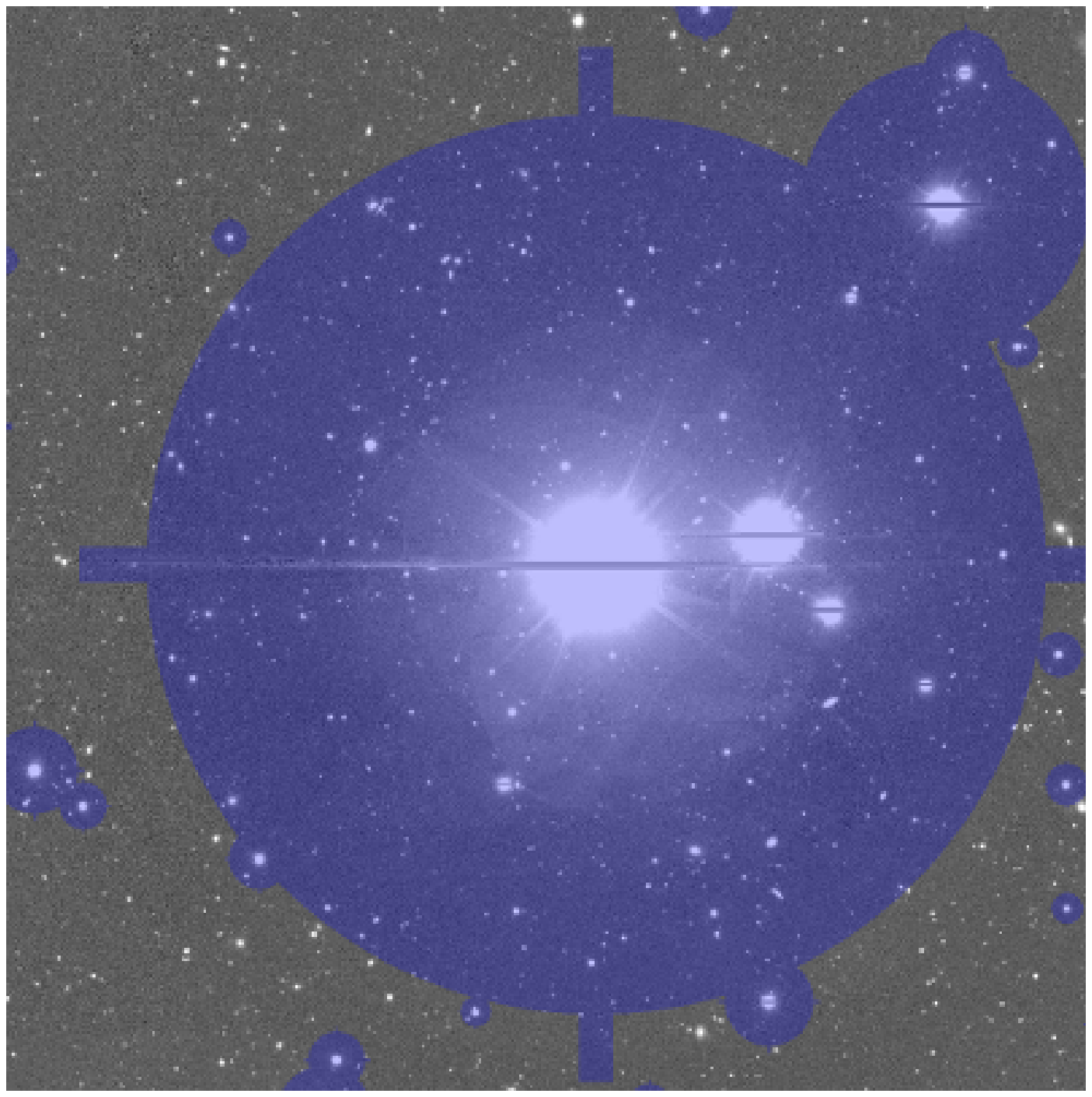}
  \end{center}
  \caption{
    Sample patch image (tract=9010, patch=4,1, $i$-band) with the bright star masks overlaid.
    The left and right panels are for old and new masks.  Note the missing mask on
    the bright star at the center ($V\sim8.7$~mag.) in the old mask.
    Also, the mask size is in general larger in the new mask.
  }
  \label{fig:brightobject}
\end{figure*}

As described in Section \ref{sec:global_sky_subtraction}, we have changed the sky
subtraction algorithm to preserve wings of bright objects.  Because of this,
masks around bright stars that indicated regions where photometry was unreliable (bright star masks; \cite{coupon18})
became too small in most cases.  In addition,
the masks used at the time of the processing were based on Gaia DR1 \citep{gaia16a,gaia16b},
and some faint stars were missing due to striping in the Gaia coverage.
There were also some bright stars that were simply missing from our catalog.
Fig.~\ref{fig:brightobject} shows an example.

All this has been largely fixed using Gaia DR2 \citep{gaia18} with a more conservative mask size.
The mask size for individual stars is determined by building an HSC source density map
for sources (with $grizy<24$) around every star in Gaia-DR2 brighter than $G=18$.  We measure the source
density profile in expanding radial annuli around the star to compute the median density
profiles as a function of star brightness, and then calculate the radius to where
the profile reached $3\sigma$ above the source density background. This is chosen as
a compromise between mask size and number of false positive peak detections.
This is performed separately for each broad-band.
In the old masks, about 12\% of the objects are masked, while the fraction increases to 20\%
in the new masks.
Note that the masks are applied to all stars brighter than $G=18$.
These new masks are being validated as of this writing.
It seems they are still not perfect but are much better than the previous version.
We plan to release the new masks no later than September 1st 2019.
We note that all the measurements are performed even for objects inside the masks.
They just have a flag bit on; {\tt \{filter\}\_pixelflags\_bright\_object[center] = 'True'}
at the database.


\subsubsection{Deblending Failure in Crowded Areas}
\label{sec:deblending_failure}

Many of the issues in PDR1 have been mitigated, but not all, and the issue of poor
photometry in crowded areas such as clusters of galaxies due to object blending still persists.
It is now a bigger problem especially in the UD-COSMOS, which goes deep enough that crowding becomes an issue
and makes the deblending even more difficult.
Some objects in UltraDeep-COSMOS have much brighter CModel magnitudes compared to PDR1, but that is likely due to deblending failures.
While an improved deblender algorithm is being developed \citep{melchior18}, the workaround for now is the same as PDR1;
PSF-matched aperture photometry on the undeblended
image (i.e., image prior to deblending) to give meaningful object colors.  The largest target seeing in PDR1 was 1.1 arcsec
and the PSF-matched photometry was not available when the original seeing was worse than that.
The largest target seeing is now increased to 1.3 arcsec,
which is the upper limit on the seeing constraint imposed in the data screening (Section \ref{sec:data_screening}).
All the coadds thus have better seeing  and the PSF-matched photometry is
always available.  Several different aperture sizes are used in the measurement,
but a small aperture (e.g., 1.5 arcsec) is recommended to avoid blending with
nearby sources.  
Note that the aperture corrections \citep{bosch18} assuming unresolved point sources
have been applied to those aperture fluxes, allowing users to obtain meaningful colors.
For extended sources, they do not give total magnitudes.

%
%

\subsubsection{Photometry in $i$ and $i2$ Combined Area}
\label{sec:photometry_in_i_and_i2}

After calibrating individual CCDs, we perform a tract-wide photometric and astrometric
calibration using multiple visits of the same field (see Section 3.2 of \cite{bosch18}).
As mentioned earlier, we combine $i$ and $i2$ data in this process.
If a tract only has either the $i$ or $i2$-band only, it is calibrated to $i$ or $i2$.
However, if a tract has a mixture of $i$ and $i2$-band visits, a processing error caused
all the visits to be calibrated to the $i$-band (which will be fixed in the next release).

The nature of this problem is somewhat complex because we apply the color term for the $i$-band
to the $i2$-band data and do the multi-visit calibration.  The difference between the $i$ and
$i2$ color terms is small, $\lesssim1$\% for most objects, except for red objects.  We apply
the color cut to avoid using late-type stars for calibration (Section \ref{sec:colorterms}),
but the photometric zero-point can be off by up to 1.5\%.  The net effect is that there is
a small offset in the zero-point between the $i2$-only region and the $i$+$i2$ combined region.
If $i$-band data dominate over $i2$, the effect is fairly small, but in regions where $i2$-band
data dominate, a clear zero-point offset is seen.

\begin{figure*}
  \begin{center}
  \includegraphics[width=16cm]{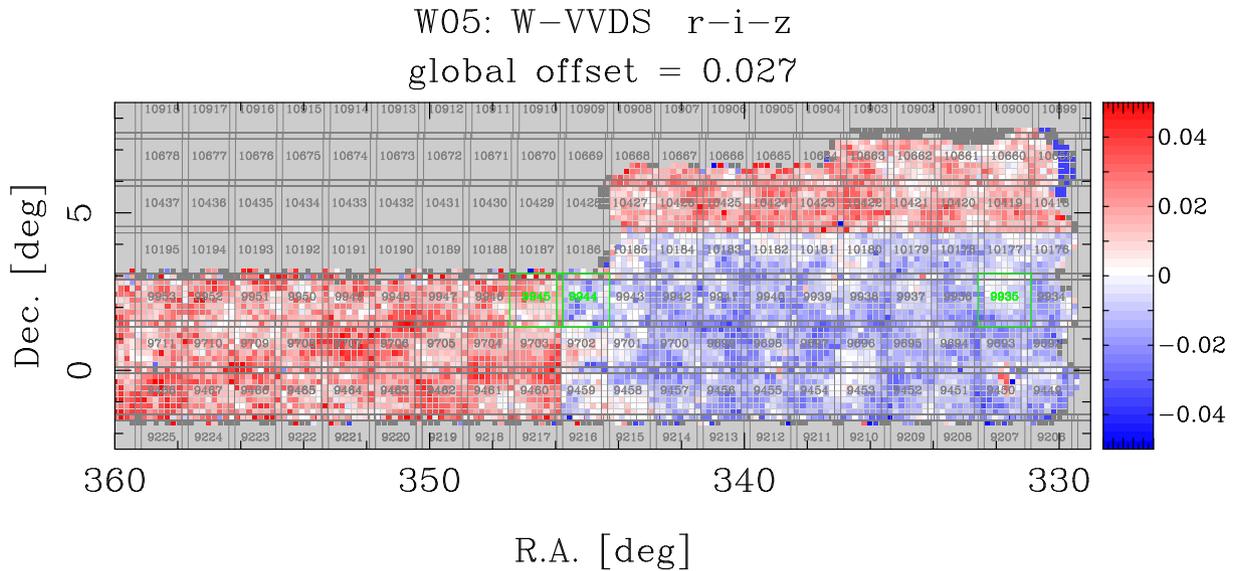}
  \end{center}
  \caption{
    Same as Fig.~\ref{fig:stellar_sequence} but for $gri$ in the Wide-VVDS field.
    The bottom-right part is calibrated to the $i$-band, while
    the rest of the area is calibrated to the $i2$-band.
    The green tracts are $i2$-only, $i$+$i2$ combined, and $i$-only regions from left to right
    and the stellar sequence in those regions are compared in Fig.~\ref{fig:vvds_stellar_sequence}.
  }
  \label{fig:vvds_color_offset}
\end{figure*}

\begin{figure}
  \begin{center}
  \includegraphics[width=8cm]{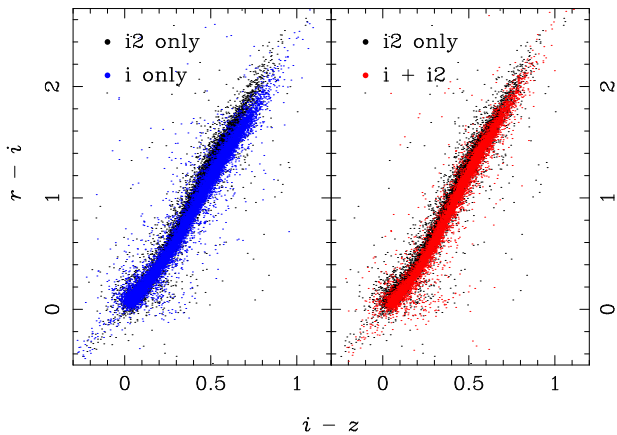}
  \end{center}
  \caption{
    $r-i$ plotted against $i-z$.  Only bright ($i<22$) stars are shown here.
    The black, blue and red points are from
    $i2$-only, $i$-only, and $i$+$i2$ combined regions indicated in Fig.~\ref{fig:vvds_color_offset}.
    The black points are common between the two panels for comparison purposes.
  }
  \label{fig:vvds_stellar_sequence}
\end{figure}

Fig.~\ref{fig:vvds_color_offset} showing the offset of the stellar sequence in $riz$
relative to the \citet{pickles98} library illustrates this problem.
The sharp tract borders at R.A.=330 deg and Dec.=4 deg are due to combination of
(1) difference between the $i$ and $i2$-bands and (2) the zero-point offset mentioned above.
Fig.~\ref{fig:vvds_stellar_sequence} compares the $riz$ stellar sequence in the $i$-band only,
$i2$-band only, and $i$+$i2$ combined regions.
First, the left panel compares $i$ and $i2$-band only regions.  The zero-points
are calibrated correctly in both bands (the stellar sequence agrees at the blue end), but
the difference in the filter transmission introduces a color difference for red stars.
In the right panel, the $i$+$i2$ combined region is actually dominated by the $i2$-band,
but there are just a few $i$-band visits that overlap with this tract and the whole tract is
calibrated to $i$.  The shape of the stellar sequence is similar to the $i2$-band only region as expected,
but because the $i$-band color term is applied to the $i2$-band data, there is a small
zero-point offset ($\Delta i=0.015$ mag).  These two effects discussed here cause
the sharp boundaries observed in Fig.~\ref{fig:vvds_color_offset}.
The $r$ and $r2$-bands have the same issue, although the effect is less severe due to
the smaller bandpass difference.

There is a database table that indicates which filter a tract is calibrated to (\code{tract_colorterm}).
There is also a database column that shows the relative fraction of $i$ and $i2$-band data that contributes
to the coadd of each object (\code{filterfraction}).  Users should refer to these tables and treat the filters
separately for applications that require accurate photometry or for objects with exotic colors.
For further analysis, the effective transmission curves discussed in Section
\ref{sec:effective_transmission_curve} will be very useful.
We may be able to provide a color term to translate $i2$ into $i$ (and vice versa) for each object to
put them on a common photometric system in a future release.


\subsubsection{Over-subtracted Scattered Light in the $y$-band}
\label{sec:over_subtracted_scattered_light_in_the_yband}

\begin{figure}
  \begin{center}
  \includegraphics[width=8cm]{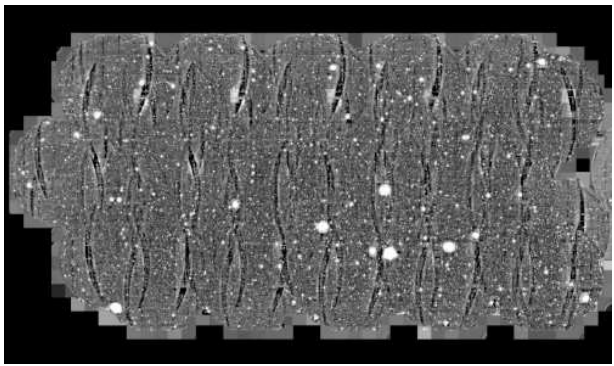}
  \end{center}
  \caption{
    $y$-band image around R.A.=157deg, approximately $9^\circ\times5^\circ$.
    The image is heavily stretched to enhance the over-subtraction feature.
  }
  \label{fig:y_oversubtracted}
\end{figure}

Due to the way the $y$-band scattered light subtraction algorithm is implemented
in the pipeline (Section \ref{sec:scattered_light_in_the_yband}), we mistakenly applied the subtraction to
all the $y$-band data taken after
the hardware fix described in Section \ref{sec:encoder_laser_shield}, in which no scattered light is observed.
This results in
an oversubtracted sky with a spatial pattern the same as the scattered light as
shown in Fig.~\ref{fig:y_oversubtracted}.  Only a small fraction of the $y$-band
data suffered from this error and the affected regions are not full-color regions (i.e.,
the regions are not yet observed in all the filters).  We thus do not expect it to be
a major issue, but users looking only at the $y$-band data should be aware of this problem.
A list of affected tracts is available at the data release site and this problem will be
fixed in a future data release.

\subsubsection{Inconsistent PSF fluxes between HSC and PS1}
\label{sec:inconsistent_fluxes}

\begin{figure*}
  \begin{center}
  \includegraphics[width=16cm]{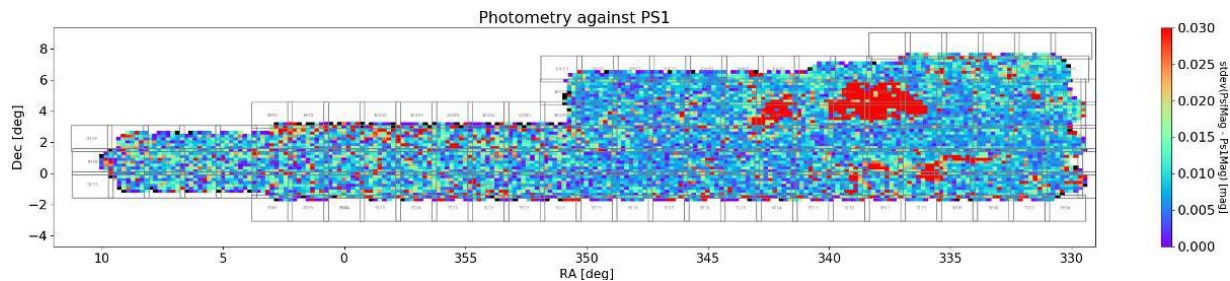}
  \end{center}
  \caption{
    Same as Fig.~\ref{fig:psfmagdiff_ps1} but for the VVDS field in the $g$-band, illustrating
    extended regions with systematic errors between HSC and PS1 PSF photometry.
  }
  \label{fig:gpsf_magdiff_vvds}
\end{figure*}

As discussed in Section \ref{sec:external_consistency}, our photometric accuracy is
good to $\sim1$\%.  However, there are regions where we observe a large scatter in
the PSF photometry between HSC and PS1.  Fig.~\ref{fig:gpsf_magdiff_vvds} shows an example,
where we see extended regions with systematic errors larger than 0.05 magnitudes.
We observe a similar scatter map when we compare with SDSS (plot not shown here) and
thus it is likely an issue with the HSC photometry.  The stellar sequence scatter seems
to follow the same spatial pattern but with a smaller scatter. 


It turns out that this is due to artifact rejection described in Section \ref{sec:artifact_rejection}.
There are 5 visits around the bad photometry area and one of them has 1.1 arcsec seeing,
while the rest have $\sim0.6$ arcsec seeing. A fraction of bright stars are clipped by the artifact
rejection algorithm in the bad seeing visit and that makes the coadd PSF incorrect;
the image itself is
based on the 4 good seeing visits, while the PSF model is constructed using all the 5 visits
(the coadd PSF does not know which visit is clipped).
This inconsistency introduced the photometry offset.  We have designed the artifact rejection
algorithm so that we do not clip real sources.  We are tracking down why a small fraction of
bright stars are actually clipped.  We will give updates at the website when we have more to report.

\subsubsection{Zero fluxes without flags in CModel}
\label{sec:zero_fluxes_without_flags_in_cmodel}


A small fraction of all the objects ($\sim1$ \%) have zero CModel fluxes with
uncertainty NaN.  This is likely a measurement failure, but the measurement flags and pixel flags
are set to \code{false} and users cannot screen them with the flags.  This issue is being
tracked down.  Users should filter out objects with CModel fluxes exactly zero
or uncertainty NaN in order not to be affected by the issue.

\subsubsection{Possible background residual}
\label{sec:possible_sky_residual}

Because we subtract the sky background on a relatively large scale, there may be a low-level sky
residual on a small scale.  A preliminary investigation seems to show a filter-dependent sky
residual at a $\sim29$ mag$/$arcsec$^2$ level.  Most sources are not affected by this level of
sky residual, but users interested in extended, low-surface brightness galaxies may want to be careful.
There is a set of useful objects in
each patch called {\it sky objects}.  The pipeline picks 100 random points in a patch outside of
object footprints and make the blank-sky measurements, just like the measurements for real objects.
These sky objects are useful for measuring background fluctuation and residual.
The sky objects are stored in the database just like real objects and they have \texttt{merge\_peak\_sky = True}.

\section{Data Access}
\label{sec:data_access}

The data can be retrieved from the data release site
where all the quality assurance plots as well as list of known issues are summarized.
As in PDR1, the release website provides only the processed data.  The raw data can be
retrieved from SMOKA\footnote{\url{https://smoka.nao.ac.jp/}}.

All the pipeline outputs are available as flat files.
There are a few online tools linked from the data release website to help users access the data they need,
such as a file search tool and an image cutout tool.
An online PSF retrieval tool allows users to retrieve the coadd PSF images at an arbitrary position on the sky.
The catalog products have been loaded to the database and users can use either
the online SQL editor or command-line tool to submit SQL queries and download the results.
The schema browser should be referred to for details of the database tables.
The online image browser, hscMap, offers a user-friendly environment to browse the massive
images.  It has many useful features (e.g., user can upload a catalog and mark objects)
and the online manual describes them.  Any questions and issues regarding the data access
should be sent to the helpdesk.

\section{Status of Collaborating Surveys}
\label{sec:status_of_collaborating_surveys}

The HSC-SSP survey has a number of collaborating surveys in other wavebands.
Here we give a brief update on two of them; a $u$-band follow-up imaging survey and
a near-IR follow-up survey. Both target the Deep/UltraDeep fields, where multiwavelength data enable a wide array of
galaxy evolution science.  

The CFHT Large Area U-band Deep Survey (CLAUDS; Sawicki et al., MNRAS submitted) has used the MegaCam imager
on the Canada-France-Hawaii 3.6m telescope to
obtain very deep U-band images that overlap the HSC-SSP Deep/UltraDeep layers.  The observations are now complete.
The new images, together with
pre-existing archival MegaCam data in some of the fields, have been processed, resampled, and stacked to match
the HSC-SSP tract/patch grid, astrometric solution, and pixel scale.  Multiband ($U+grizy$) photometry is carried out
using SExtractor \citep{bertin96}
and an adaptation of \code{hscPipe} that can handle these CFHT $U$-band images.
The CLAUDS data cover 18.60~deg$^2$ with median seeing of FWHM=0.92" and to a median depth of $U = 27.1$ AB
(5$\sigma$ in 2" apertures); selected areas in the COSMOS and SXDS fields that total 1.36~deg$^2$ reach a median depth of
$U=27.7$ AB (5$\sigma$ in 2" apertures).  Altogether, the CLAUDS images represent the equivalent of 113 classical-mode
CFHT nights and are the deepest $U$-band data ever taken over this combination of depth and area.  The combined
CLAUDS and HSC-SSP datasets enable many science investigations by significantly enhancing photometric redshift
performance and allowing the selection of $z\sim 3$ Lyman Break Galaxies and quasars. Several science projects are
already underway with this combined dataset, and the CLAUDS team anticipates releasing these deep $U$ images and data
products (including CLAUDS+HSC-SSP $U+grizy$ catalogs based on HSC-SSP PDR2 data) to the public in 2020.


Turning to near-infrared data, the Deep and UltraDeep fields overlap with some of the major
near-infrared imaging surveys such as the Deep eXtragalactic Survey of the UKIRT Infrared
Deep Sky Survey (UKIDSS/DXS; \cite{kim11}), Ultra Deep Survey with the VISTA Telescope
(UltraVISTA; \cite{mccracken12}), VISTA Deep Extragalactic Observations Survey (VIDEO;
\cite{jarvis13}).  These surveys, however, do not fully cover the Deep fields,
and Deep UKIRT Near-infrared Steward Survey (DUNES$^2$; Egami et al., in
preparation) is filling the missing part.

DUNES$^2$ has made excellent progress; the data acquisition has been essentially complete
with a total observing time of about 270 hours on UKIRT.
DUNES$^2$ is similar to UKIDSS/DXS in terms of depth, and covers the four flanking fields of E-COSMOS
($J$\,$\sim$\,23.6, $H$\,$\sim$\,23.2, $K$\,$\sim$\,23.2 mag at $5\sigma$ within 2 arcsec aperture; 3.0 deg$^2$ in total) and
DEEP2-3 field ($J$\,$\sim$\,23.3, $K$\,$\sim$\,23.1 mag; 4.5 deg$^2$).  DUNES$^2$ also obtained $H$-band data for the central
0.9\arcmin$\times$1.7\arcmin\ region of ELAIS-N1 ($H\sim23.2$ mag; 1.5 deg$^2$), which was missing from
the UKIDSS/DXS survey.  The data are being processed, and our plan is to use the HSC photometry pipeline to
produce fully band-merged source catalogs covering from the $U$ to $K$ bands, following the methodology developed
for the CLAUDS $U$-band data.   We anticipate that such catalog products as well as images
will be made available publicly at the time of the HSC-SSP final data release (DR3).
We also note that a further follow-up near-IR imaging survey, DeepCos, designed to bring the majority (5.6 deg$^2$) of
the DUNES$^2$ footprint to the depth of VIDEO ($J=24.5$ and $K=23.5$), is currently on-going at UKIRT
(Lin et al. in prep.).

\section{Summary and Future Data Releases}
\label{sec:summary}

The data from 174 nights of HSC-SSP are now publicly available.  The data are of high quality
and should enable a wide range of scientific explorations.  However, there are known issues 
with the data and we advise users to review the issue list (Section \ref{sec:known_issues}) before using the data.
We also ask users to acknowledge HSC-SSP; the sample acknowledgment text is given at the data release site.
In addition, the HSC technical papers given in Table \ref{tab:references} should be referred to where appropriate.
The pipeline is developed as part of LSST and therefore the appropriate LSST papers should also be referenced:
\citet{ivezic19}, and \citet{juric17}.
We have calibrated our data against the public Pan-STARRS data.  We would like to encourage
users to reference Pan-STARRS as well:
\citet{chambers16}, \citet{schlafly12}, \citet{tonry12}, and \citet{magnier13}.

Looking towards the future,
our baseline plan is to make the next major data release (PDR3) in two years, but our observations
suffered from bad weather in winter 2017-2018 as well as from earthquakes due to the increased
volcanic activity at Kilauea in summer 2018.  There was an additional hiatus due to a telescope problem
in September-October 2018.  The survey has been significantly delayed due to these problems and
this may affect our data release plan.  We will give updates on our website in due course.

\begin{table}[htbp]
  \begin{center}
    \begin{tabular}{cccc}
      \hline
      Subject                 &  Paper \\
      \hline
      Survey design           & \citet{aihara18b}\\
      Public Data Release 1   & \citet{aihara18a}\\
      Public Data Release 2   & this paper\\
      Camera system           & \citet{miyazaki18}\\
      Camera dewar            & \citet{komiyama18}\\
      Filters                 & \citet{kawanomoto18}\\
      Processing pipeline     & \citet{bosch18}\\
      Onsite reduction system & \citet{furusawa18}\\
      SynPipe                 & \citet{huang18}\\
      Bright object masks     & \citet{coupon18}\\
      Photometric redshifts   & \citet{tanaka18}\\
      Lensing shape catalog   & \citet{mandelbaum18}\\
      \hline 
    \end{tabular}
  \end{center}
  \caption{
    List of HSC technical papers.
  }
  \label{tab:references}
\end{table}

\section*{Acknowledgments}
The Hyper Suprime-Cam (HSC) collaboration includes the astronomical communities of Japan and Taiwan,
and Princeton University.  The HSC instrumentation and software were developed by the National
Astronomical Observatory of Japan (NAOJ), the Kavli Institute for the Physics and Mathematics of
the Universe (Kavli IPMU), the University of Tokyo, the High Energy Accelerator Research Organization (KEK),
the Academia Sinica Institute for Astronomy and Astrophysics in Taiwan (ASIAA), and Princeton University.
Funding was contributed by the FIRST program from Japanese Cabinet Office, the Ministry of Education,
Culture, Sports, Science and Technology (MEXT), the Japan Society for the Promotion of Science (JSPS),
Japan Science and Technology Agency  (JST),  the Toray Science  Foundation, NAOJ, Kavli IPMU, KEK, ASIAA,
and Princeton University.

This paper makes use of software developed for the Large Synoptic Survey Telescope. We thank the LSST
Project for making their code available as free software at http://dm.lsst.org.

The Pan-STARRS1 Surveys (PS1) and the PS1 public science archive have been made possible through contributions by the Institute for Astronomy, the University of Hawaii, the Pan-STARRS Project Office, the Max-Planck Society and its participating institutes, the Max Planck Institute for Astronomy, Heidelberg and the Max Planck Institute for Extraterrestrial Physics, Garching, The Johns Hopkins University, Durham University, the University of Edinburgh, the Queen's University Belfast, the Harvard-Smithsonian Center for Astrophysics, the Las Cumbres Observatory Global Telescope Network Incorporated, the National Central University of Taiwan, the Space Telescope Science Institute, the National Aeronautics and Space Administration under Grant No. NNX08AR22G issued through the Planetary Science Division of the NASA Science Mission Directorate, the National Science Foundation Grant No. AST-1238877, the University of Maryland, Eotvos Lorand University (ELTE), the Los Alamos National Laboratory, and the Gordon and Betty Moore Foundation.

This paper is based on data collected at the Subaru Telescope and retrieved from the HSC data archive system, which is operated by Subaru Telescope and Astronomy Data Center at National Astronomical Observatory of Japan.  Data analysis was in part carried out with the cooperation of Center for Computational Astrophysics, National Astronomical Observatory of Japan.

We thank the anonymous referee for a thoughtful report, which helped improve the paper.
This work is also based on zCOSMOS observations carried out using the Very Large Telescope at the ESO Paranal Observatory under Programme ID: LP175.A-0839, on observations taken by the 3D-HST Treasury Program (GO 12177 and 12328) with the NASA/ESA HST, which is operated by the Association of Universities for Research in Astronomy, Inc., under NASA contract NAS5-26555, on data from the VIMOS VLT Deep Survey, obtained from the VVDS database operated by Cesam, Laboratoire d'Astrophysique de Marseille, France, on data from the VIMOS Public Extragalactic Redshift Survey (VIPERS). VIPERS has been performed using the ESO Very Large Telescope, under the "Large Programme" 182.A-0886. The participating institutions and funding agencies are listed at http://vipers.inaf.it.  Funding for SDSS-III has been provided by the Alfred P. Sloan Foundation, the Participating Institutions, the National Science Foundation, and the U.S. Department of Energy Office of Science. The SDSS-III web site is http://www.sdss3.org/.  SDSS-III is managed by the Astrophysical Research Consortium for the Participating Institutions of the SDSS-III Collaboration including the University of Arizona, the Brazilian Participation Group, Brookhaven National Laboratory, Carnegie Mellon University, University of Florida, the French Participation Group, the German Participation Group, Harvard University, the Instituto de Astrofisica de Canarias, the Michigan State/Notre Dame/JINA Participation Group, Johns Hopkins University, Lawrence Berkeley National Laboratory, Max Planck Institute for Astrophysics, Max Planck Institute for Extraterrestrial Physics, New Mexico State University, New York University, Ohio State University, Pennsylvania State University, University of Portsmouth, Princeton University, the Spanish Participation Group, University of Tokyo, University of Utah, Vanderbilt University, University of Virginia, University of Washington, and Yale University.  GAMA is a joint European-Australasian project based around a spectroscopic campaign using the Anglo-Australian Telescope. The GAMA input catalogue is based on data taken from the Sloan Digital Sky Survey and the UKIRT Infrared Deep Sky Survey. Complementary imaging of the GAMA regions is being obtained by a number of independent survey programmes including GALEX MIS, VST KiDS, VISTA VIKING, WISE, Herschel-ATLAS, GMRT and ASKAP providing UV to radio coverage. GAMA is funded by the STFC (UK), the ARC (Australia), the AAO, and the participating institutions. The GAMA website is http://www.gama-survey.org/.  Funding for the DEEP2 Galaxy Redshift Survey has been provided by NSF grants AST-95-09298, AST-0071048, AST-0507428, and AST-0507483 as well as NASA LTSA grant NNG04GC89G.  Funding for PRIMUS is provided by NSF (AST-0607701, AST-0908246, AST-0908442, AST-0908354) and NASA (Spitzer-1356708, 08-ADP08-0019, NNX09AC95G). Funding for the DEEP3 Galaxy Redshift Survey has been
provided by NSF grants AST-0808133, AST-0807630, and AST-0806732.
This work is in part supported by MEXT Grant-in-Aid for Scientific Research on Innovative 
Areas (No.~15H05887, 15H05892, 15H05893).

\bibliographystyle{apj}
\bibliography{references}

\end{document}